\def\LP#1{}
\def\AJ#1{}
\def\LP#1{{\color{blue} \textsl{\small $^{[LP]}[$#1$]$}}} 
\def\AJ#1{{\color{red} \textsl{\small $^{[AJ]}[$#1$]$}}} 

\documentclass{aa}  

\usepackage{color, colortbl}
\usepackage{graphicx}
\usepackage{tikz}
\usepackage{float}
\usepackage{subcaption}
\usepackage{multicol}
\usepackage{longtable}
\usepackage{placeins}
\usepackage{hyperref}
\usepackage{ulem}
\usepackage{txfonts}

\begin{document}

   \title{Resolving the mass transfer
in the symbiotic recurrent nova
T~Coronæ~Borealis\thanks{Based on observations made with the Mercator Telescope, operated
on the island of La Palma by the Flemish Community, at the Spanish
Observatorio del Roque de los Muchachos of the Instituto de Astrofísica
de Canarias}}

 \author{L. Planquart
          \inst{1}\fnmsep\thanks{Research fellow, FNRS, Belgium.}
          \and
          A. Jorissen \inst{1}
          \and
        H. Van Winckel \inst{2}
          }

   \institute{Institut d’Astronomie et d’Astrophysique, Université Libre de Bruxelles, CP 226, Boulevard du Triomphe, 1050 Brussels,
Belgium\\
              \email{lea.planquart@ulb.be}
         \and
             Institute of Astronomy, KU Leuven, Celestijnenlaan 200D, B-3001 Leuven, Belgium \\
             }

   \date{Received 31/10/2024; accepted 29/12/2024}

 
  \abstract
   { T~Coronæ~Borealis (T~CrB) is a symbiotic recurrent nova with an 80-year recurrence interval whose next eruption is imminent.}
    {We aim to resolve the accretion mechanism of the binary system governing the mass transfer during its super-active phase.}
    {Using phase-resolved high-resolution spectroscopy, we analyze the zoo of spectral-line profiles arising from the symbiotic activity. We perform Doppler tomography of selected emission lines to resolve the system's gaseous components and their different velocity regimes.}
    {We find evidence of enhanced accretion through Roche lobe overflow during the super-active phase, as traced by the oxygen, helium, and hydrogen lines. The accretion disc is found to be fully viscously evolved and extends up to its maximal radius. 
    By mapping the kinematics of lines probing different excitation energies, we can identify distinct interaction sites. These include the bright spot at the stream impact on the accretion disc outer radius, the irradiation at the red-giant facing side, the stream-disc overflow, the accretion disk wind, and an expanding nebula.
    The nebula emerged at the rise of the super-active phase and underwent an acceleration phase of about five years. The temporal evolution of the lines supports the scenario where the departure from quiescence started in the disc, likely triggered by a disc instability similar to what occurs in dwarf novae outburst, leading to an increased mass accretion and causing important irradiation of the giant that has further enhanced the mass-transfer rate during the super-active phase. }
    {Symbiotic recurrent nova, such as T~CrB, are governed by similar mass-transfer mechanisms as found in cataclysmic variables despite their different orbital properties (longer orbital periods imposing larger accretion discs) and evolutionary pathways. }


   \maketitle
%

\section{Introduction}
\label{sec:introduction}

Cataclysmic variables (CVs) are interacting binaries composed of a white dwarf (WD) primary accreting material from a Roche-filling later-type donor. Depending on their photometric behavior, they are subdivided into classical novae (CN, one recorded eruption with an amplitude typically larger than $6$\,mag), recurrent novae (RN, more than one eruption recorded), dwarf novae (DN, with frequent and less intense outbursts\footnote{In the paper, the term 'eruption' refers to a nova explosion caused by a thermonuclear runaway at the WD surface. The term 'outburst' refers to outburst events seen in dwarf novae that are triggered by an instability in the accretion disc.} of 2$-$5\,mag) and nova-like systems (NL, regrouping all the non-eruptive CVs) \citep{CV_book_cvs..book.....W}.


T Coronæ Borealis (T~CrB) is a prototypical symbiotic recurrent nova. It has two previously reported eruptions in 1866 and 1946 \citep{CV_book_cvs..book.....W}, and possibly two others in 1217 and 1787 \citep{previous_nova_explosiion_2023JHA....54..436S}. The system is composed of an evolved M giant in a 227-d orbit \citep{Kraft_1958ApJ...127..625K} with a WD companion, whose nature is inferred from the ultraviolet emission lines (such as \ion{He}{ii}, \ion{N}{v}, \ion{C}{iv}) and hot continuum \citep{IUE_data_1992ApJ...393..289S}. 
The T~CrB recurrent eruptions are assumed to be driven by a thermonuclear runaway at the surface of a WD close to the Chandrasekhar mass (\citealt{Kato_massive_wd_1999PASJ...51..525K}, \citealt{mass_wd_132_2018ApJ...860..110S}). The $Gaia$ parallax of the system is 1.09~$\pm$~0.03\,mas \citep{gaia_2023A&A...674A...1G}, translating to a nominal distance of $895_{-23}^{+22}$\,pc \citep{distance_bailer_jones_2021AJ....161..147B}. 
Each of the two recorded nova eruptions displays a peculiar photometric behavior: a decade-long phase of increased flux (up to 1~mag in the optical) precedes the eruption \citep{TCrB_prediction_outburst_2023MNRAS.524.3146S}, and a secondary maximum appears $~$5 months after the nova explosion \citep{Stanford_1949ApJ...109...81S}. The origin of the secondary maximum is still unclear: it could be caused by the irradiation of the red giant by a tilted accretion disc (AD) surrounding the WD \citep{tilted_disc_second_bump_1999ApJ...517L..47H} or by the cooling WD whose AD has been destroyed after the explosion \citep{cooling_WD_2023RNAAS...7..251M}.  

After the last eruption, the system remained in the quiescence phase until 2015, when the system suddenly brightened up. The mass accretion rate abruptly increased by about 20 times its value in quiescence \citep{TCrB_prediction_outburst_2023MNRAS.524.3146S}, reaching $2-6 \times10^{-8}~ \rm M_\odot$/yr (\citealt{Luna_2018A&A...619A..61L}, \citealt{accretion_superactive_2023A&A...680L..18Z}, \citealt{Xray_ejections_2024MNRAS.532.1421T}, \citealt{TCrB_prediction_outburst_2023MNRAS.524.3146S}), and several changes across the wavelength range were reported.
In the X-ray range, the hard X-ray (E~$\geq$~2\,keV) flux decreased, while a soft component (E~$\leq$~0.6\,keV) appeared, implying that the boundary layer between the WD surface and the inner region of the AD switched to an optically thick regime (\citealt{Luna_2018A&A...619A..61L, Luna_BL_2019ApJ...880...94L}). In the optical, the appearance and strengthening of emission lines and blue continuum were reported \citep{super_active_phase_2016NewA...47....7M}. The $B$-band flux increased by 0.9~mag and deviated from its ellipsoidal variations (see Fig.\ref{fig:Blightcurve} and \citealt{TCrB_prediction_outburst_2023MNRAS.524.3146S}). The rising blue continuum is associated with an effective temperature of $9400\pm500$\,K attributed to the hot component \citep{accretion_superactive_2023A&A...680L..18Z}. A flux increase was also detected in the radio band, probing bremsstrahlung emission of ionized material surrounding the system \citep{SAP_radiorange_2019ApJ...884....8L}.
All those multi-wavelength indicators reflect that the system has entered a so-called super-active phase (SAP; \citealt{super_active_phase_2016NewA...47....7M}) during which a significant amount of mass was accreted onto the compact star ($\sim 2\times10^{-7}~\rm M_\odot$; \citealt{accretion_superactive_2023A&A...680L..18Z}). It suggests that most of the mass needed to trigger the thermonuclear runaway could have been accreted during the recent SAP \citep{Luna_2020ApJ...902L..14L}. The SAP of T~CrB stopped in 2023, hinting at its imminent nova eruption, predicted around 2025.5$\pm$1.3 \citep{TCrB_prediction_outburst_2023MNRAS.524.3146S}. 
While the SAP seems to be an essential precursor in the recurrent nova eruption, the underlying binary interaction still remains to be understood. 
T~CrB represents a key system at the crossroads of different families of interacting binaries: symbiotic system, recurrent nova, and cataclysmic variable. It exhibits some unique characteristics, such as the pre-eruption SAP and the post-eruption second maximum. It is, therefore, a crucial object to constrain the accretion physics of long-period interacting binaries and to understand the binary evolution of potential progenitors of Type Ia supernovae (\citealt{Hachisu_TcrB_outburst_2001ApJ...558..323H}, \citealt{SN1a_2019A&A...622A..35L}).

In this paper, we shed new light on the binary interaction during the SAP, by spectrally resolving the accretion flows that drive the mass transfer of the symbiotic system. 
We derived the orbital elements during the SAP (Sect.~\ref{sec:system_properties}). Then, we analyze the phase-dependent behaviors of spectral lines and assess their locations within the system using Doppler tomography (Sect.~\ref{sec:spectral_monitoring}). Next, we evaluate the temporal evolution of selected lines during the SAP (Sect.~\ref{subsect:timing_lines}). Finally, we discuss the possible origin of the system recent behavior and its link with the expected nova eruption (Sect.~\ref{sec:discussion}). We end up with a summary of our main findings (Sect.~\ref{sec:conclusions}).

\section{Observations and orbital model}
In this section, we retrieve the system orbital parameters from the spectroscopic monitoring and derive its orbital model. 
\label{sec:system_properties}
T CrB has been monitored with the HERMES spectrograph mounted on the 1.2m Mercator telescope at La Palma \citep{HERMES}. The high-resolution mode was used, providing a resolution of $R=86\,000$ over the wavelength range from 3800 to 9000~\AA. A hundred spectra were taken on different dates between January 2011 and June 2023, among which 83 were obtained during the SAP (2015$-$2023),  with a median exposure time of 900 seconds. The observation log can be found in Table \ref{tab:RV_data}.

\subsection{Spectroscopic orbit}
\label{subsec:spectroscopic_orbit}
The Keplerian orbit was retrieved from the radial velocity (RV) curve using a $\chi^2$ minimization.
The RV values were obtained by cross-correlating the object spectrum with an MIII template  -- compatible with the spectral type estimates: M4.5 \citep{Spect_type_TCrB_B1999A&AS..137..473M} or M3III \citep{super_active_phase_2016NewA...47....7M} -- and fitting a Gaussian function to the mean line profile, as represented by the resulting cross-correlation function. 
In the fitting process, the orbital period was initially set to the maximum of the Lomb-Scargle periodogram. 

\begin{figure}[ht]
    \centering
    \includegraphics[width = 0.5\textwidth]{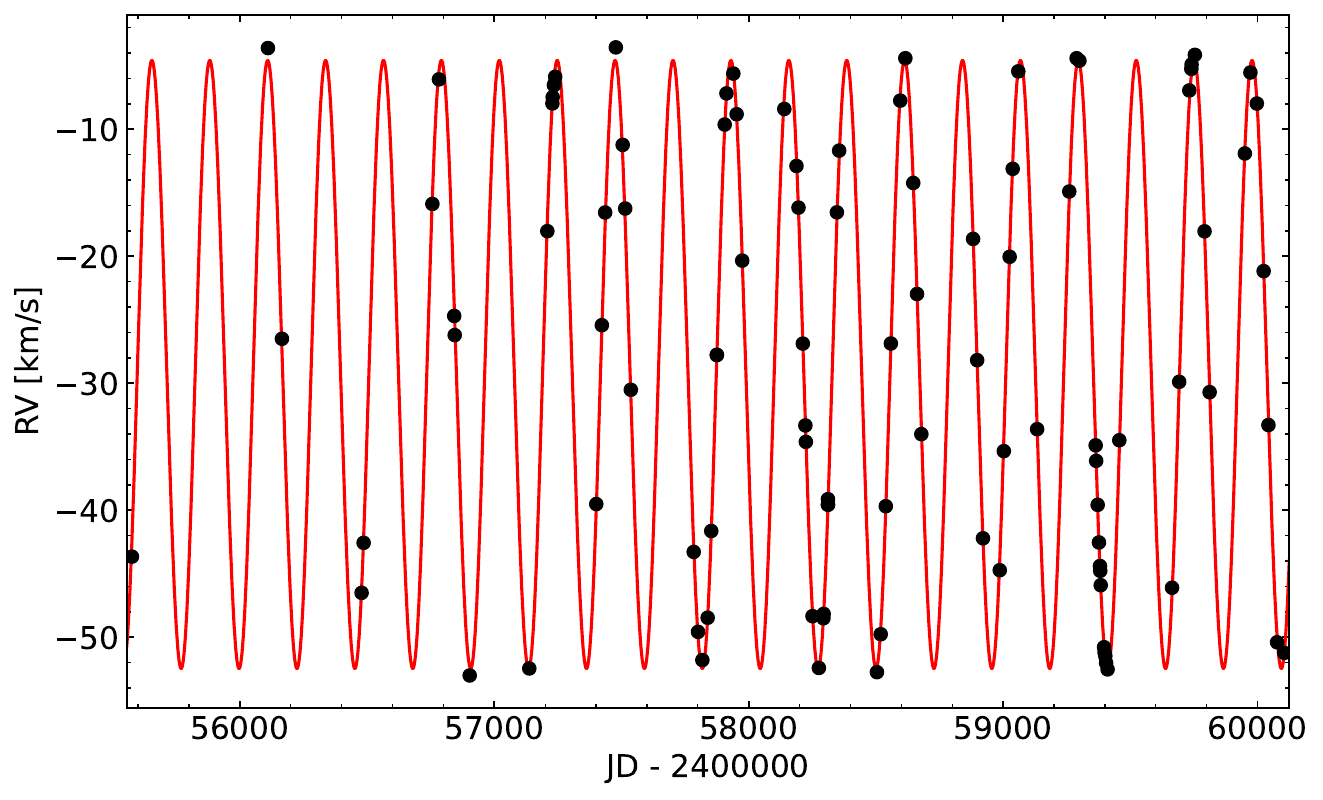}
    \includegraphics[width = 0.5\textwidth]{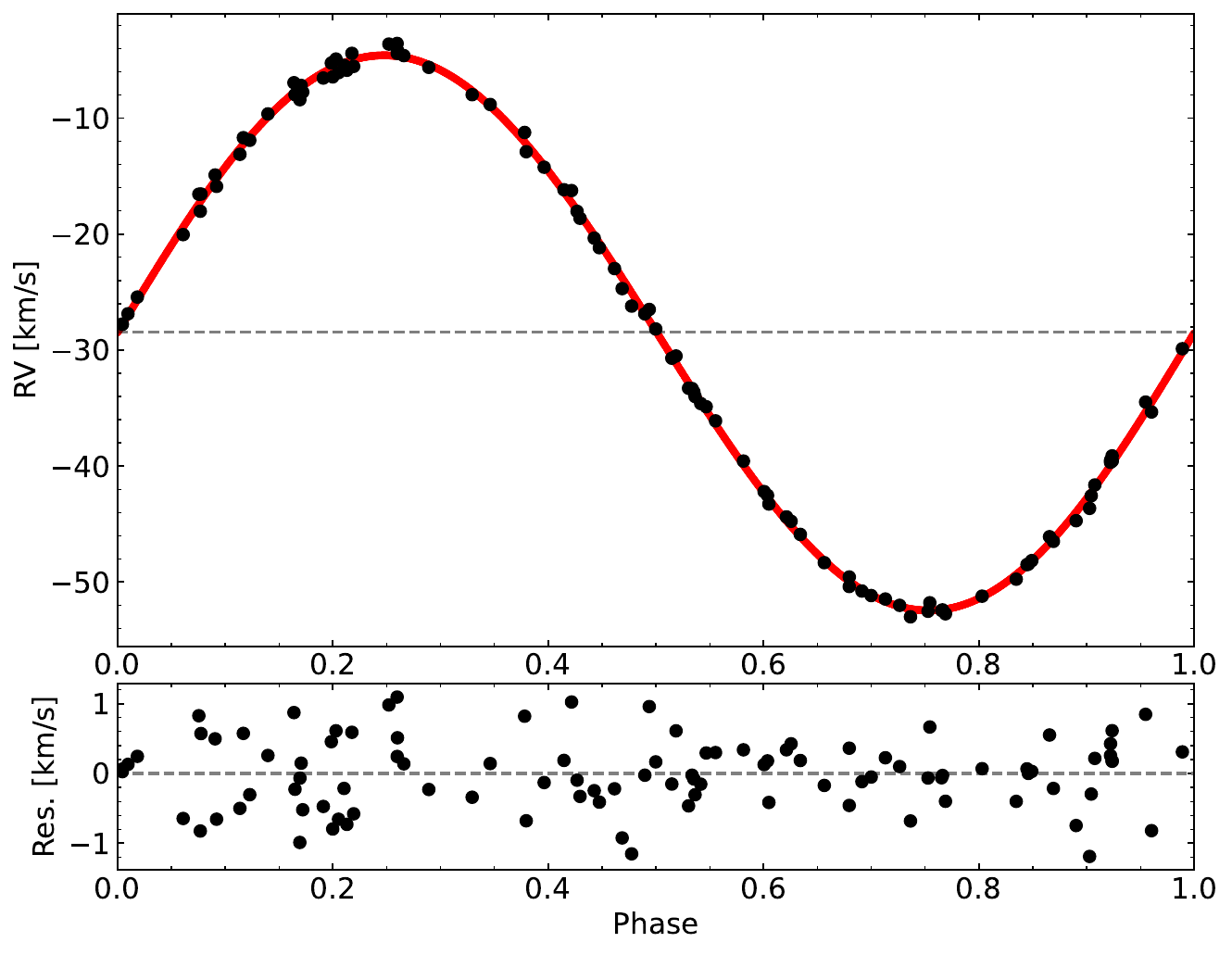}
    \caption{Radial velocities of T CrB. The error bars are plotted as well, but they are smaller than the symbol size of the data points. Top: RV curve as a function of time. Middle: Phase-folded orbit of T~CrB. The full red line is the Keplerian orbit. Bottom: Residual values.}
    \label{fig:orbit}
\end{figure}

The obtained parameters are listed in Table \ref{tab:orbital_parameters} and compared with the values obtained from infrared RV measurements by \cite{infrared_rv_2000AJ....119.1375F}. Their good agreement, except for $V_\gamma$ related to the offset between the instrumental systems, implies that no significant change in the orbital period has occurred between the quiescent phase and the SAP. The RV curve is shown in Fig.~\ref{fig:orbit}.
The standard deviation of the O-C residuals ($\sigma_{OC}$ = 0.5~km/s) is larger than the instrumental error (0.07\,km/s;  \citealt{error_hermes_rv_2016A&A...586A.158J}) and is probably caused by the intrinsic jitter observed for M giants \citep{famaey_2009A&A...498..627F}.

\begin{table}[t]
    \centering
        \caption{Orbital parameters for the giant in the T CrB system.}
    \begin{tabular}{lrrr}
    \hline
    \hline
        Parameters & 2013-22 (1) &1997-99 (2)\\
        \hline
        $P$ [d]& 227.58 $\pm$ 0.03 & 227.57 $\pm$ 0.01\\
        $e$ & 0.009 $\pm$ 0.003 &0.0\\
        $K_1$ [km/s]& 23.94 $\pm$ 0.13& 23.89 $\pm$ 0.17\\
        $V_\gamma$ [km/s]& -28.47 $\pm$ 0.05& -27.79 $\pm$ 0.13\\
        $a_1 {\rm sin} i$ [au]& 0.500 $\pm$ 0.003& 0.488 $\pm$ 0.004\\
        $f(m)$ [$\rm M_{\odot}$]&  0.324 $\pm$ 0.005 & 0.322 $\pm$ 0.007\\
        $T_0$ [HJD] & 2\,455\,825.44 $\pm$ 0.08& 2\,447\,861.73 $ \pm$ 0.27\\
        $T'_0$ [HJD]  & $-$& 2\,455\,826.68 $ \pm$ 9.45\\
        $R^2$ & 99.91\% & $-$\\
            \hline
    \end{tabular}
    \tablebib{(1) This work; (2) \cite{infrared_rv_2000AJ....119.1375F}}
\tablefoot{$P$ is the orbital period, $e$ the eccentricity, $K_1$ the semi-amplitude of the giant's RV curve, $V_\gamma$ the center-of-mass velocity, $a_1$ the semi-major axis of the orbit of the giant around the center-of-mass of the system, $i$ the inclination, $f(m)$ the mass function, $T_0$ is the epoch of inferior conjunction,  $T'_0$ is the epoch of inferior conjunction of (2) extrapolated to the epoch of (1), and $R^2$ is the regression coefficient. }
    \label{tab:orbital_parameters}
\end{table}

\begin{table}[t]
    \centering
        \caption{Derived orbital parameters of the system.  }
    \begin{tabular}{lrrr}
    \hline
    \hline
        Parameters & Value & Formula\\
        \hline
        $i$ [$^\circ$] & 65 $\pm$ 5 & Fixed, (1)\\
        $M_{wd}$ [$\rm M_{\odot}$]& 1.32 $\pm$ 0.10 & Fixed, (2)\\
        $q$ &  $0.74 \pm 0.18$ & derived from $f(m), i, M_{wd}$\\
        $M_{RG}$ [$\rm M_{\odot}$]& $0.98 \pm 0.31$& $q\times M_{wd}$\\
        $a$ [au]&  $0.96 \pm 0.06$ & $(1+q) \times a_1$\\
        $b_1$ [au] & $0.51 \pm 0.01$& $a\times(0.500- 0.227\log(q))$\\
        $R_{RG}$ [au]&  $0.34\pm 0.04$&  $a\times (0.38 + 0.20\log(q))$, (3)\\
        $r_{circ}$ [au] & $0.13 \pm 0.01$&  $b_1^4/(P^2M_{wd})$ \\
        $r_{t}$ [au]  & $0.33\pm0.02$ & $0.6a/(1+q)$, (4)\\

            \hline
    \end{tabular}
    \tablebib{(1) \citealt{RV_stanishev_2004A&A...415..609S}; (2) \citealt{mass_wd_132_2018ApJ...860..110S}; (3) \citealt{Paczynski_1971ARA&A...9..183P}; (4) \citealt{CV_book_cvs..book.....W}.}
\tablefoot{For the definition of the parameters, see text.}
    \label{tab:orbital_parameters_2}
\end{table}
\subsection{Orbital model and disc size}
\label{subsec:orbital_model}
As the RV curve only gives partial information on some orbital parameters and does not give access to the individual masses, additional constraints should come from complementary studies, such as the analyses of photometric data.
From the ellipsoidal variations observed in the light curve, the orbital inclination is constrained between 60$^\circ$ and 70$^\circ$ (\citealt{1998MNRAS.296...77B}, \citealt{RV_stanishev_2004A&A...415..609S}). 
From amplitude and decline time modeling of the nova eruption, the WD mass estimates lie between $1.32\pm0.10~\rm M_\odot$ \citep{mass_wd_132_2018ApJ...860..110S} and $1.37\pm0.01~\rm M_\odot$ \citep{Kato_massive_wd_1999PASJ...51..525K}. 
In this paper, we adopt $M_{WD} = 1.32\pm0.10~\rm M_\odot$ \citep{mass_wd_132_2018ApJ...860..110S}, $i=65^\circ\pm5^\circ$ \citep{RV_stanishev_2004A&A...415..609S}, leading to $M_{RG}=1.0 \pm 0.3 \rm ~M_\odot$ and $q=0.74\pm0.15$. The mass ratio is later confirmed by the \ion{He}{ii} monitoring (see Sect.~\ref{subsubsec:helium_ionized}).

As the system is semi-detached, the donor radius is equal to its Roche lobe, $R_{RG}$, and the mass transfer is, therefore, governed by Roche lobe overflow, where the equations of the restricted three-body problem describe the accretion flow's analytical path. The flow originates from the $L_1$ point at the surface of the Roche-filling donor, at a distance $b_1$ from the WD, and follows a ballistic trajectory until it reaches the outer radius of the AD \citep{hot_spot_location_1972MNRAS.160...15W}. The location of the stream impact within the outer radius defines the bright spot (sometimes referred to as the hotspot). Two boundaries fix the extension of the AD. The circularization radius, $r_{circ}$, sets the lower bound to the outer radius and corresponds to the radius of the circular Keplerian orbit having the same angular momentum as that carried by the matter coming from the $L_1$ point. The tidal-truncation radius, $r_{t}$, defines the upper bound of the AD-outer radius \citep{accretion_book_2002apa..book.....F}. After the stream hits the AD at its outer rim,  the flow is deflected and spread vertically, creating a disc overflow that can veil the inner disc from the observer for $i\geq 65^\circ$ \citep{veiling_2001MNRAS.322..499K}.
Tab.~\ref{tab:orbital_parameters_2} lists the derived parameters that scale the binary system. 


\section{Orbital-phase variation of the spectral lines}
\label{sec:spectral_monitoring}
While the spectrum of the red giant largely dominates the overall flux in the visible and near-infrared ranges, the bluest part ($\lambda <4500$~\AA) is blended by the rising blue continuum associated with the hot component during the SAP. 
Table \ref{tab:spectral_lines} lists the principal spectral line signatures indicative of symbiotic behavior observed during the SAP. Those are detected in emission or absorption, either with a velocity or line strength variation. 

\begin{figure*}[ht]

    \includegraphics[width = 0.3\textwidth]{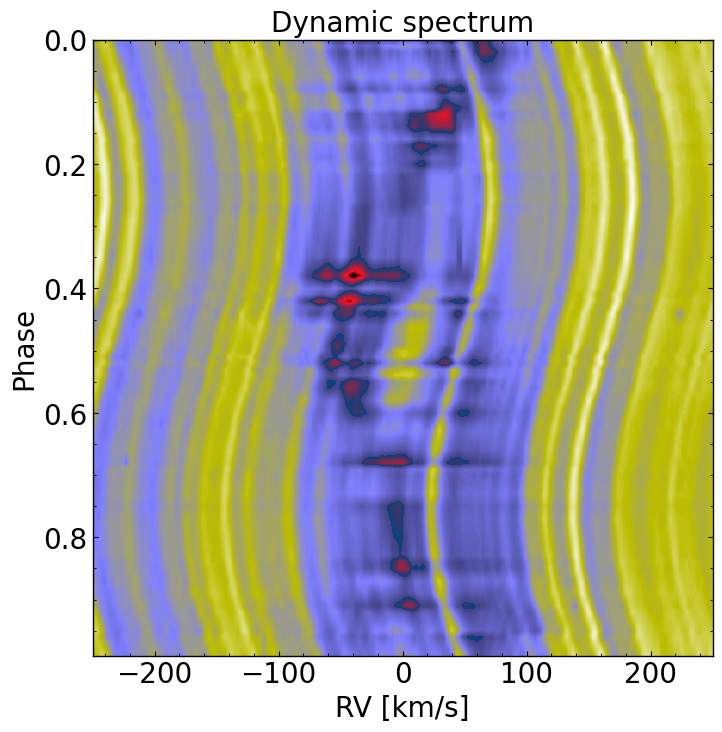}
    \hfill
    \includegraphics[width = 0.3\textwidth]{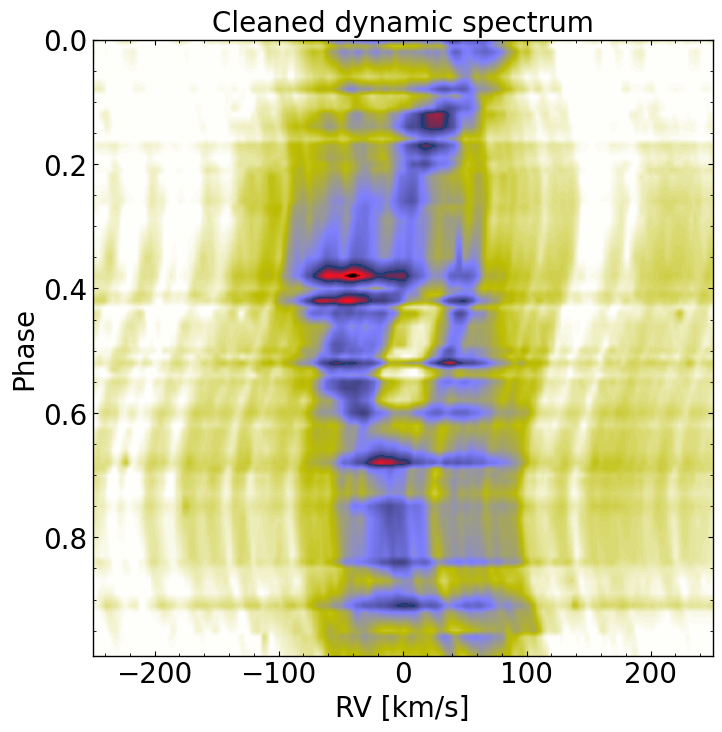}
    \hfill
    \includegraphics[width = 0.32\textwidth]{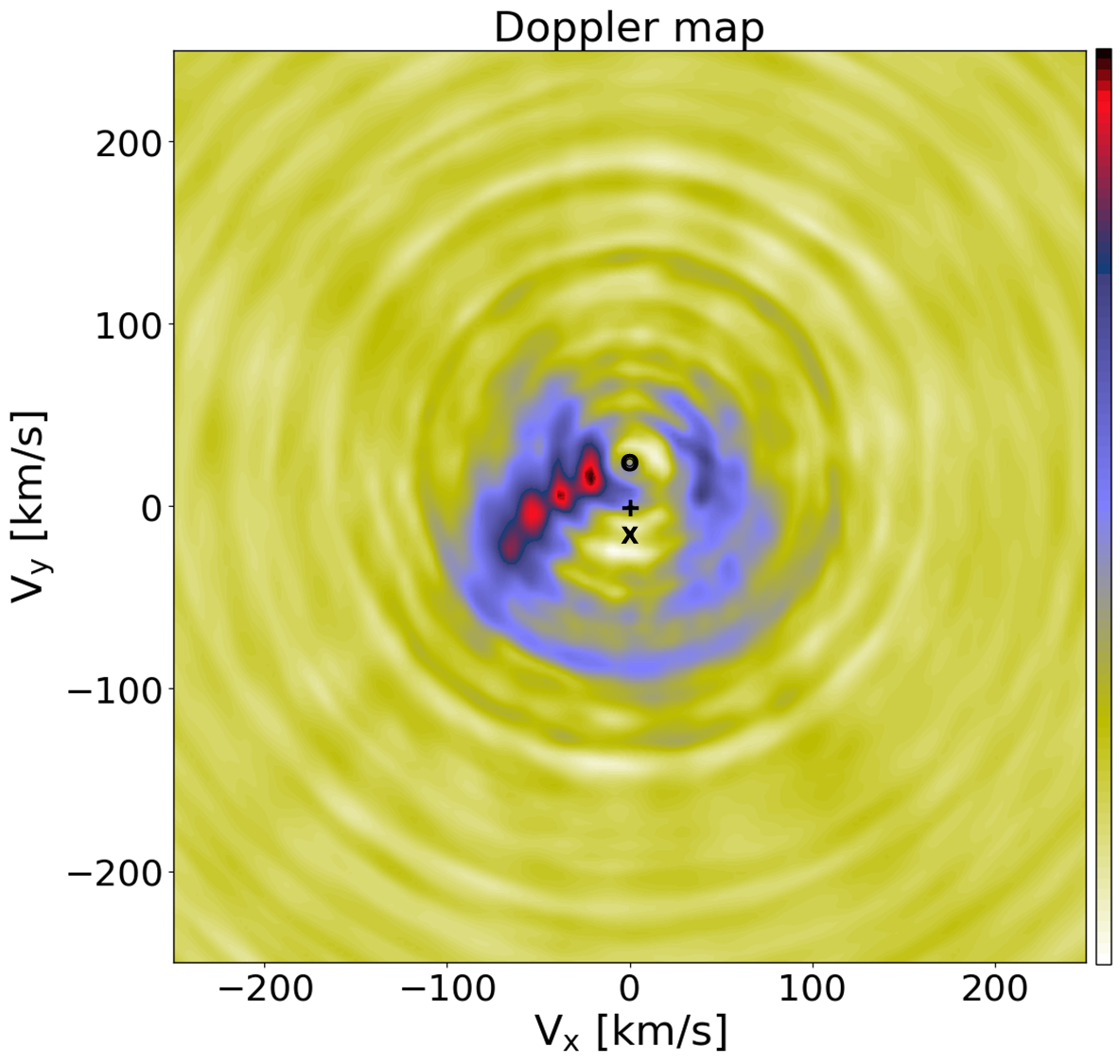}

    \includegraphics[width = 0.64\textwidth]{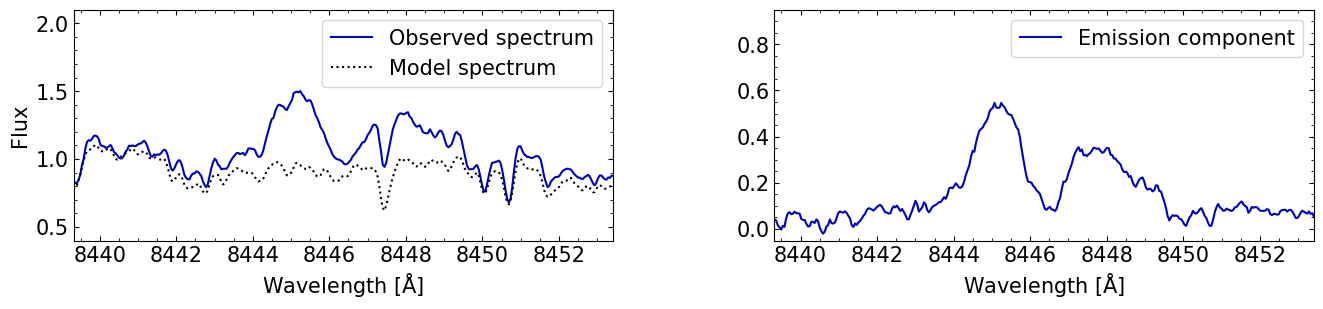}\hfill \hfill    
    
    \caption{Computation of the Doppler map for the \ion{O}{i} line. The dynamic spectrum from the observation (left), the dynamic spectrum after removing the giant contribution (middle), and the Doppler map (right). The color of the dynamic spectra represents the pseudo-continuum-normalized fluxes taken as the median value of the spectral window. The color of the Doppler maps corresponds to arbitrary units of emission intensity. On the Doppler map, the position of the donor is given by an empty circle, that of the accretor by a cross, and the center of mass by the plus sign. The bottom panels show the spectral profile before (left) and after (right) the subtraction of the giant contribution for a spectrum at $\phi = 0.55$.}
    \label{fig:doppler_explaination}
\end{figure*}

In this section, we analyze the dependence of the spectral lines on the orbital phase. First, we describe the emission lines detected on top of the giant spectrum and assess their location through Doppler tomography and their evolution during the SAP. Then, we report on those absorption line profiles whose behavior shows an unusual change associated with the orbital motion. 
Doppler tomography consists of spatially mapping the emission intensity in the velocity space from time-resolved spectroscopic observations. \cite{Doppler_tomo_1988MNRAS.235..269M} first introduced this technique to interpret the complex S-wave line variations found in the spectra of CVs, as it allows us to directly compare the observation with the model of an accreting binary translated into its velocity coordinates lying in the orbital plane ($V_x$, $V_y$). In such representation, the signature of an accretion disc is seen inside-out, as the inner disc appears as a ring at large velocities while the outer edge probes lower velocities. 
The Doppler maps are computed using the filtered back-projection method of the \textsc{iradon} function from the \textsc{scikit-image} \textsc{Python3} package. The Doppler maps are performed on the cleaned dynamic spectrum, obtained by subtracting, for each observed spectrum, a model spectrum of the giant Doppler-shifted to the appropriate phase. The result of the cleaning and Doppler mapping is illustrated in Fig.~\ref{fig:doppler_explaination} for the line \ion{O}{i} line at 8446\AA. 
Appendix~\ref{appendix:spetra_and_Doppler_maps} displays the individual phase-folded interpolated spectra of the spectral lines described below and, for the emission lines, their corresponding Doppler maps.

\begin{table}[ht]
\caption{Inventory of the spectral lines associated with the symbiotic activity.}
    \centering
    \begin{tabular}{lrrl}
    \hline
    \hline
        Element & $\rm \lambda$ [\AA] &  $\chi_u$ [eV]& Comment \\
        \hline
        \ion{H}{i}   &8750.46 &13.50 &weak absorption\\
        & $^*$6562.85 & 12.09 &\\
        & $^*$4861.34 &  12.75&\\ 
        & 4340.47&  13.06&\\ 
        & 4101.75&  13.22&blended by \ion{N}{iii}\\
        & 3970.07& 13.32&blended by \ion{Ca}{ii} \\
        & 3889.06& 13.39&blended by \ion{He}{i} \\

        \ion{He}{i} & 7065.18& 22.72&$\rm ^3 S_1$\\
        & $^*$6678.15&  23.08& $\rm ^1 S_0$ \\
        & $^*$5875.64& 23.08&$\rm ^3 S_1$ \\
        & 5015.68&  23.07& $\rm ^1 S_0$ \\
        & 4921.93& 23.74&$\rm ^1 S_0$ \\
        & $^*$4713.15&  23.59& absorption  \\
        & 4471.48& 23.74&absorption \\
        
        & 4026.19& 24.04&absorption\\
        & $^*$3964.73&  23.74&absorption  \\
        & 3888.65& 23.01& blended by \ion{H}{i}\\
        \ion{He}{ii}  & $^*$4685.70&  51.02 &\\
        
        \ion{N}{iii} & 4634.14 & 33.13 & \ion{He}{ii} fluorescence\\
         & $^*$4640.64 & 33.13 & \ion{He}{ii} fluorescence\\
        & 4097.33 &  30.46& blended by \ion{H}{i}\\
        
        \ion{O}{i} & $^*$8446.36& 10.99 & Ly$\beta$ fluorescence\\      
        & $^*$7775.39& 10.74& absorption\\
        & $^*$7774.17& 10.74& absorption\\
        & $^*$7771.94&  10.74& absorption\\
        
        $[$\ion{O}{iii}$]$ & $^*$5006.84 & 2.51&\\ 
        & 4958.91 & 2.51&\\ 
        & 4363.21 &  5.35&\\
        
        $[$\ion{Ne}{iii}$]$& $^*$3868.76 & 3.20& \\
        
        
        \ion{Ca}{ii} & 8662.14   &3.12&blended emission\\
        & 8498.02  & 3.15&blended emission\\
        & 8542.09 &3.15&blended emission\\

         \hline
    \end{tabular}
    
    \tablefoot{Lines marked with $*$ in front of their wavelengths $\lambda$ have their phase-folded spectra displayed in Appendix \ref{fig:spectra_interpol}. $\chi_u$ is the upper energy level taken from the NIST database (\url{https://physics.nist.gov/asd}). The atomic configuration of the Helium lines, singlet $\rm ^1 S_0$ or triplet $\rm ^3 S_1$, in emission, is also given. }
    \label{tab:spectral_lines}
\end{table}

\subsection{Neutral atom transitions}
The neutral helium and hydrogen lines are all found in emission on top of the giant spectrum (for $\lambda > 4500$\,\AA), together with the \ion{O}{i} transition at 8446\,\AA.  

The \ion{He}{i} lines are found in two flavors: the orthohelium (triplet state, e.g., 5875\,\AA~transition) and the parahelium (singlet state, e.g., 6678\,\AA~line). The variation of the emission profile with the orbital phase differs despite similar excitation energies (see Table~\ref{tab:spectral_lines}), hence similar excitation temperatures ($T>20\,000$~K; \citealt{lithium_2020AJ....159..231W}): the \ion{He}{i} 6678\,\AA~line shows a narrow "S-wave" emission pattern while the \ion{He}{i} 5875\,\AA~line is found to exhibit a more complex profile made of a varying broad emission blended with a blueshifted absorption component (Fig.~\ref{fig:spectra_interpol}). The dual behavior of para and ortho \ion{He}{i} lines can serve as a probe of different density regions, as already reported by \cite{CV_Helium_2003AJ....126.1472K} for the old nova Q~Cygni. 
It can be understood in terms of atomic considerations: in the orthohelium configuration, the lowest state is metastable and located 19.8\,eV above the ground state. As a result, the levels are more likely to be overpopulated due to collisional excitation than their parahelium counterparts, where radiative de-excitation to the ground state can occur, contrary to the situation prevailing for orthohelium. For high-density regions, we can expect parahelium transitions to show a pure emission profile and orthohelium transitions to display an absorption component instead. 

Similar considerations can be made when comparing H$\alpha$ with the 8446\AA\ \ion{O}{i} line. The H$\alpha$ line (and the other Balmer transitions) exhibits a broad ($\pm 150$~km/s) emission, more pronounced around phase 0.5 and nearly absent at phase 0, where a blueshifted additional absorption component is superimposed on the emission profile, leading to a P-Cygni profile. On the other hand, the \ion{O}{i} line exhibits a similar S-wave profile (see Fig.~\ref{fig:doppler_explaination}) as the \ion{He}{i} 6678~\AA\ line, with an additional absorption component 
originating from the part of the giant photosphere facing the disc,
around phase 0.5 (see Sect.~\ref{subsubsec:from_the_donor}). The \ion{O}{i} 8446 \AA~line is known to be a fluorescence line excited by the \ion{H}{i}~Ly$\beta$ photons. Hence, both the \ion{O}{i}/Ly$\beta$ and H$\alpha$ transitions depend on the population of the third level of \ion{H}{i}.
But while \ion{O}{i}/Ly$\beta$ can only be excited by radiative recombination through fluorescence mechanism, the H$\alpha$ line profile in symbiotic binaries is expected to bear the signature of several processes: the core is formed in a narrow transition zone near the ionization front \citep{H_alpha_core_1997A&A...319..166S} while its wings are believed to be dominated by Raman scattering of Ly$\beta$ photons \citep{Raman_yes_2000ApJ...541L..25L} or by an optically-thin stellar wind from the WD \citep{Raman_no_2006A&A...457.1003S}.

\begin{figure*}[t]
    \sidecaption
    \includegraphics[width =12cm]{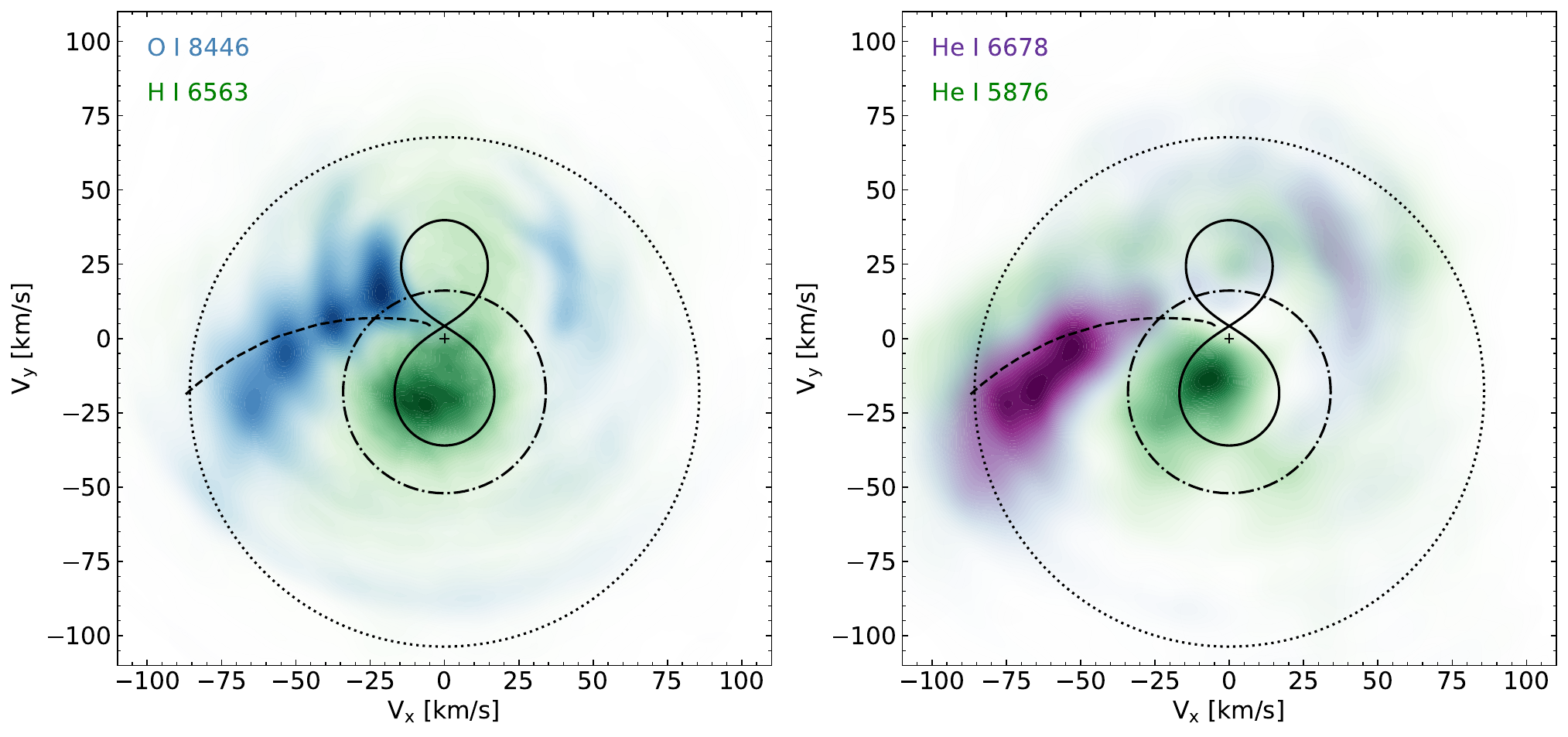}
    \caption{Combined Doppler tomograms. The Roche lobe (black solid line), the center of mass (black plus), the ballistic flow velocity (dashed line), the Keplerian velocities of the AD at $r_{circ}$ (dotted line) and $r_{t}$ (dash-dotted line) are marked. Left: Hydrogen-based transition. Right: Helium-based transition. }
    \label{fig:combined_Doppler_maps}
\end{figure*}

To further assess the location of formation of the emission lines, their respective Doppler tomograms are superimposed in Fig.~\ref{fig:combined_Doppler_maps} and compared with the velocity path of the accretion stream and the Keplerian velocity at the AD outer radii [$r_{circ}$, $r_{t}$] derived in Sect.~\ref{subsec:orbital_model}. In \ion{O}{i} and \ion{He}{I}, the emission pattern forms a clear arc-shaped spot located in the negative $V_x$ region between the Keplerian velocities of the two accretion radii. It can be interpreted as originating from the bright spot, where the stream impacts the material from the AD. The \ion{O}{i} emission is located at the intersection of the stream with the lower velocity circle associated with the tidal truncation radius.
In contrast, the \ion{He}{i} emission is enclosed between those two circles of Keplerian velocity. Hence, we can conclude that the \ion{He}{i} 6678\AA\,line is excited closer to the circularization radius than \ion{O}{i} that is already excited at the disc's outer edge. By resolving the bright spot in the velocity space, we confirm the presence of a large AD around the WD extending up to its tidal-truncation limit, filling almost 70\% of the Roche surface at the WD.
In contrast, the Balmer H$\alpha$ and the \ion{He}{i} 5875~\AA\ lines have a shallower emission pattern peaking on the WD velocity and an absence of emission at the position of the bright spot. We suggest these lines arise instead from the accretion wind surrounding the AD or the WD.

\subsection{Forbidden transitions}
\label{subsubsec:forbideen_transition}
Forbidden lines of [\ion{O}{iii}] and [\ion{Ne}{iii}] are found in emission, with a pattern composed of a double emission profile centered near the center-of-mass velocity, creating a ring-like structure in the Doppler maps (Fig.~\ref{fig:doppler_maps}). The spectral profile is invariant against the orbital phase, indicating that the lines -- collisionally excited and probing low-density regions -- are formed in the nebula surrounding the system.  From the detection of the auroral line [\ion{O}{III}] at 4363~\AA, we obtained a ratio $\log ((I_{5007}+ I_{4959})/I_{4363}) = 0.8\pm0.1$,  implying an electron density larger than $\sim10^6~\rm cm^{-3}$ (Fig.~9 from \citealt{OIII_diagnostic_2023MNRAS.521.4750H}).

\begin{figure}[t]
    \centering
    \includegraphics[width=\linewidth]{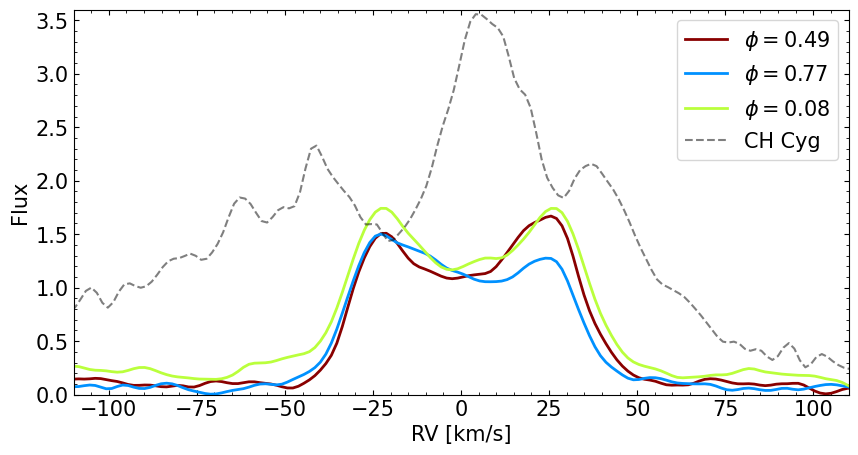}
    \caption{Spectral profiles of the [\ion{O}{III}] line at 5007~\AA\ at different orbital phases of the same cycle. The profiles are obtained after subtracting the giant contribution. The $y$-axis is the pseudo-continuum-normalized flux obtained as the median value of the spectral window. One spectrum of the symbiotic system CH~Cyg is shown for comparison.}
    \label{fig:spectral_profile_5007}
\end{figure}

Considering that the [\ion{O}{iii}] transition in the spatially-integrated spectra exhibits a symmetrical double-peaked line profile (Fig.~\ref{fig:spectral_profile_5007}), two opposite geometries for the emitting region could be envisioned: either the double-peaked emission is the signature of circumbinary material forming an equatorial ring in the orbital plane and expanding radially with a de-projected velocity of $25/\sin(65^\circ)\approx28~\rm km/s$, or the double-peaked emission is associated with a bipolar jet, launched perpendicular to the orbital plane (e.g., from the AD), with a de-projected velocity of $25/\cos(65^\circ)\approx60~\rm km/s$. 
The nebula of symbiotic stars can also be found to exhibit a more complex morphology, composed of knotty polar filaments and an equatorial torus -- e.g., the optical nebula of CH~Cyg (\citealt{CHCyg_1996MNRAS.278..542T}, \citealt{CHCyg_2001ApJ...560..912C}) -- but it should result in a more complex emission profile (see Fig.~\ref{fig:spectral_profile_5007}).

Given that these lines are formed in low-density regions and that the [\ion{O}{iii}] line at 5007\AA\, has been found in the polar region of some symbiotic systems -- see R~Aqr \citep{RAqr_OIII_2018A&A...612A.118L} or EG~And \citep{O5007_EG_And_2021A&A...646A.116S} --, we favor the explanation where the forbidden lines are formed in the polar region of the system. 
Coincidentally, bipolar ejections were also suggested by \cite{Xray_ejections_2024MNRAS.532.1421T} to explain the appearance of the soft X-ray component in the T~CrB spectrum during the SAP. In their analysis, the soft X-ray emission results from the shock of the matter ejection perpendicular to the AD with the circumstellar medium surrounding the system. From their modeled plasma temperature for the soft emission ($1.7 -9.7 \times10^5$\,K), they estimate a jet velocity, $v$, of 110$-$200\,km/s, a factor two higher than our measurements in the optical. If the line traces the propagation of a blast wave, the post-shock plasma temperature follows a $v \propto \sqrt{T}$ dependency; hence, a lower velocity by a factor two implies temperature reduction by a factor four. The forbidden line of [\ion{O}{iii}] would correspond to a temperature region of about 45\,000~K.

\subsection{Ionized helium transitions}
\label{subsubsec:helium_ionized}

The high excitation emission lines of \ion{He}{ii} 4686\AA~and \ion{N}{iii} (Bowen fluorescence of \ion{He}{ii} at 303~\AA) are found in emission during the SAP and can be attributed to a hot ($T>60\,000$~K) region in the system \citep{lithium_2020AJ....159..231W} 

\begin{figure}[t]
    \centering
    \includegraphics[width=\linewidth]{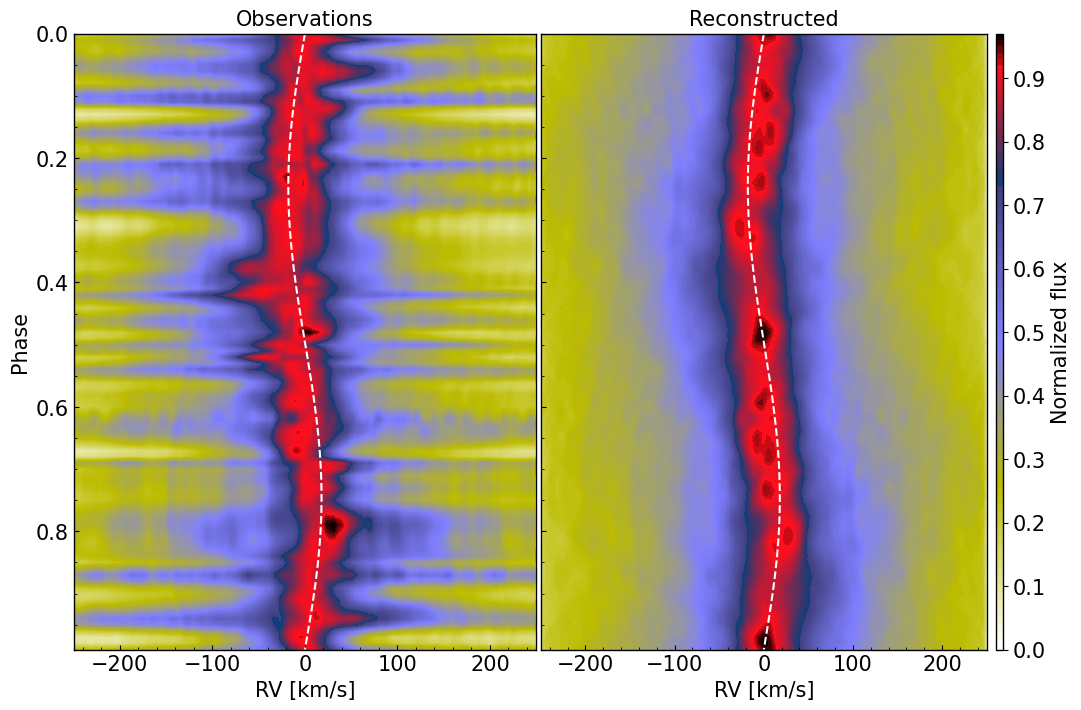}
    \caption{\ion{He}{ii} orbital variation. Left: observation after subtraction of the giant contribution and second normalization. Right: reconstructed dynamic spectrum. The motion of the white dwarf is shown with a white dashed line.  }
    \label{fig:he_II}
\end{figure}

Their emission profile exhibits a narrow single-peaked emission profile with a small RV-variation ($\leq 50$~km/s) through the orbital cycle. These lines exhibit a strong intensity modulation independent of the orbital phase (but that could be correlated with hard X-ray variation, according to \citealt{flickering_2016MNRAS.462.2695I}), that leads to a flickering aspect of the dynamic phase-resolved spectrum (Fig.~\ref{fig:spectra_interpol}). 
To assess the location of the emitting source, disregarding this additional amplitude modulation, each spectrum (after subtraction of the giant contribution) was re-normalized to set its maximum intensity to unity. The cleaned phase-resolved spectra and the reconstruction from Doppler mapping are shown in Fig.~\ref{fig:he_II}. The emission profile follows a clear sinusoidal modulation in phase with the WD motion, with a velocity offset of $-6\pm3$km/s to the center-of-mass velocity. The semi-amplitude of the peak maximum is $16\pm2$~km/s, comparable to the WD predicted semi-amplitude of $K_2 = q \times K_1 = 18\pm4$\,km/s (from Tables \ref{tab:orbital_parameters} and \ref{tab:orbital_parameters_2}), confirming the previous estimates of the accretor mass and mass ratio. Such a pattern results in a broad emission spot around the WD velocity in the Doppler maps (Fig.~\ref{fig:doppler_maps}). 
Thus, the ionized emission lines (\ion{He}{ii} and its fluorescent transitions) originate from the region close to WD. They could result from reprocessed X-ray radiation at the boundary layer between the AD and the WD or come directly from the inner part of the AD wind from near the WD, as found and modeled in some high-inclination nova-like systems, as the result of a high mass-transfer rate (see \citealt{HeII_as_wind_1994MNRAS.267..153H}, \citealt{SWSex_1996ApJ...471..949H}).
\subsection{Absorption lines from the donor}
\label{subsubsec:from_the_donor}

\begin{figure*}[t]
    \centering
    \includegraphics[width = 0.3\textwidth]{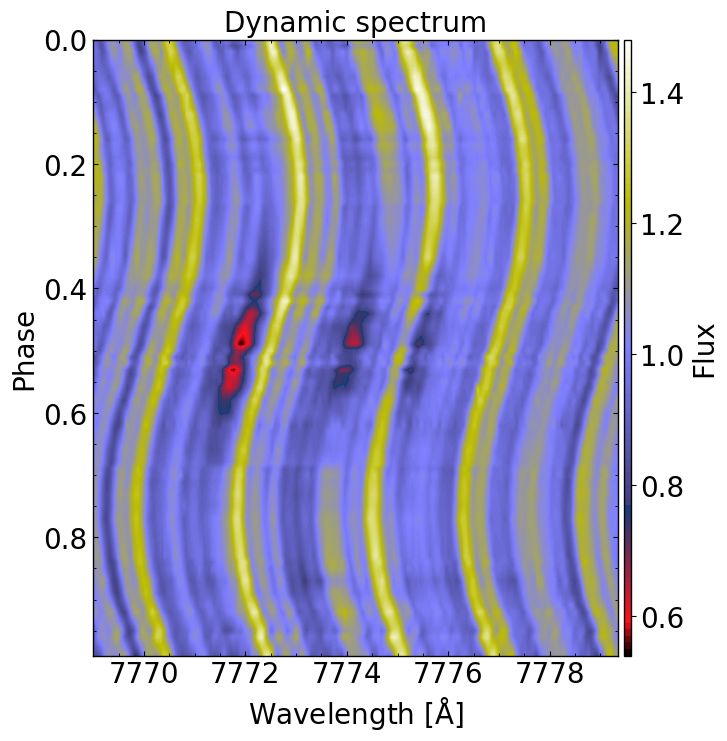}
    \includegraphics[width = 0.5\textwidth]{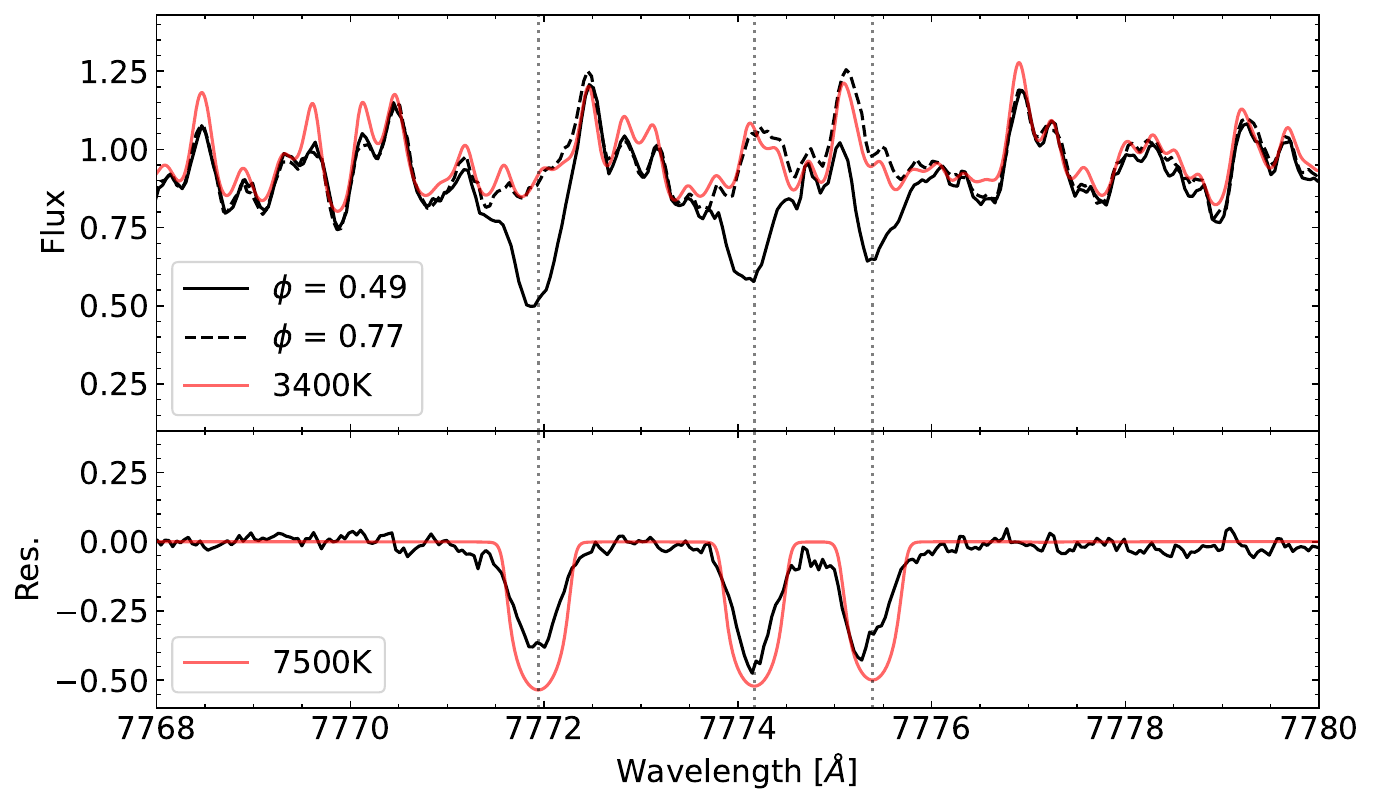}
    \caption{\ion{O}{i} triplet lines. Left: dynamic spectrum. Middle: T~CrB spectra at phases 0.49 (black solid line) and 0.77 (dashed line), and a synthetic spectrum at 3400~K (red line). The bottom panel displays the difference spectrum between phases 0.49 and 0.77 (black line) and a synthetic spectrum at 7500~K. Right panel: surface mapping of the residual absorption in the difference spectrum. The orbital phase marks the corresponding viewing angle.}
    \label{fig:o_triplet_profile_variation}
    
\end{figure*}

The absorption lines of the donor dominate the optical spectrum.
However, in the near-infrared part of the spectrum, the depth of the calcium and oxygen triplets varies according to the orbital phase. 
The calcium triplet exhibits a shallower absorption profile from phases 0.3 to 0.5, attributed to an emission blend from the symbiotic activity. 
In the case of the oxygen triplet at 7772\,\AA, the absorption lines abruptly deepen from phases 0.45 to 0.55. This is illustrated in Fig.~\ref{fig:o_triplet_profile_variation}. The left panel shows the dynamical spectrum of T~CrB centered on the \ion{O}{i} triplet, while the middle panel illustrates the spectral profile at two different orbital phases. Comparison with synthetic spectra obtained using \textsc{marcs} model atmospheres and \textsc{Turbospectrum} LTE \citep{turbospectrum_2012ascl.soft05004P} is superimposed. The giant parameters used are taken from  \cite{metallicity_abundance_2023MNRAS.526..918G}: $T_{\rm eff}$= 3400~K, $\log g=0.5$, [Fe/H] = 0.35, [O/H] = 0.1.

Since the \ion{O}{i} lines follow the orbital motion, they must form within the Roche lobe of the giant. Figure~\ref{fig:o_triplet_profile_variation} (right panel) illustrates the phase dependence of the \ion{O}{i} line at 7772~\AA, which experiences a 40\% absorption-depth variation with the orbital phase\footnote{A similar dependence is found for the Paschen line at 8750.46~\AA, but the line depth only reaches 10\%.}. It is clear that the extra absorption originates from the side facing the WD and could hint at irradiation or deformation effects near the $L_1$ nozzle\footnote{It should be noted that similar behavior is seen in the classical symbiotic EG~And (Planquart et al., in prep.)}. 
Apart from the \ion{O}{i} triplet line, the general shape of the giant spectrum does not change significantly with the orbital phase: the TiO lines that dominate the spectrum undergo a slight modulation interpreted as an increase of $\rm T_{eff}$ by $+80$~K, likely to be due to irradiation \citep{super_active_phase_2016NewA...47....7M}. But the additional absorption depth seen in the \ion{O}{i} triplet lines cannot be explained by a change of the surface integrated $\rm T_{eff}$ -- as an increase of +80~K does not lead to prominent absorption lines -- or $\log g$ -- as it would affect other lines as well. Instead, it could hint at a localized irradiated spot on the deformed side of the giant. The difference spectrum between orbital phases 0.49 and 0.77, plotted in the bottom left panel of Fig.~\ref{fig:o_triplet_profile_variation}, is compared with a synthetic spectrum of $\rm T_{eff}=7500$\, K, as the \ion{O}{i} triplet is found in stellar atmospheres with $\rm T_{eff}$>5000\, K \citep{OI_models_A-K_2018PASJ...70....8T}. While beyond the scope of the current paper, synthetic models considering the 3D effects at the irradiated side of the deformed giant could be foreseen to derive the temperature gradient on the stellar surface. 

\subsection{Absorption lines following the accretor}
\label{subsubsec:from_the_accretor}
\begin{figure}[t]
    \centering

    \includegraphics[width = 0.24\textwidth]{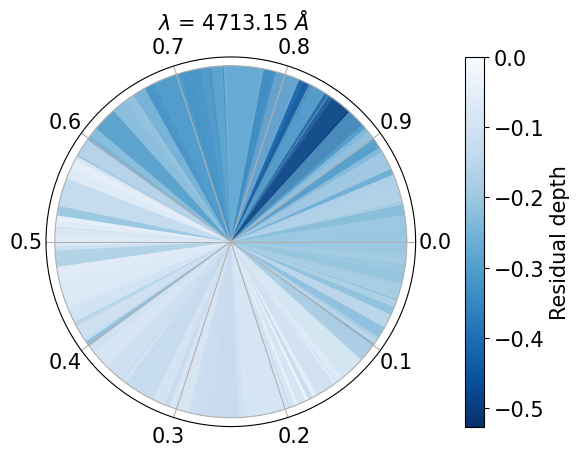}
    \includegraphics[width = 0.24\textwidth]{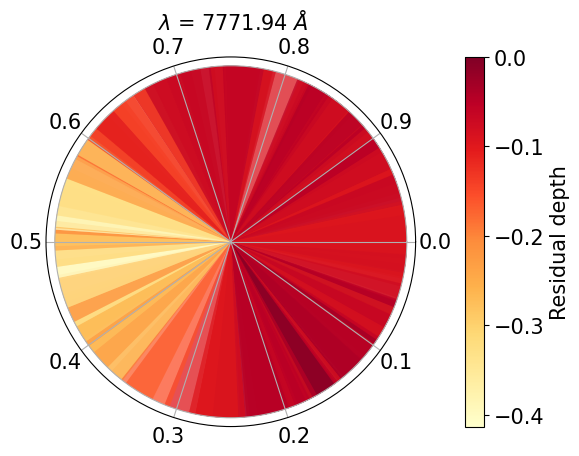} 
    \caption{Surface mapping of the intensity of the absorption lines. The orbital phase (with 0.5 corresponding to the superior conjunction when the companion is located in front of the giant) are marked at  
    their corresponding viewing angle. Left: \ion{He}{i} line following the accretor showing that the absorption deepens at the stream-disc impact ($\phi = 0.6-0.9$). Right: \ion{O}{i} line following the donor showing that the absorption deepens on the side facing the AD ($\phi=0.4-0.6$).}
    \label{fig:He_abs_profile_variation}
    
\end{figure}

\ion{He}{i} lines at 4713.15~\AA\ and below are found in absorption and anti-phase with the giant motion. Their detection in the spectrum is associated with the rising blue continuum. But while their variation is in phase with the WD motion, their amplitude exhibits higher velocities, especially around $\phi=0.8$: up to 60~km/s compared with the WD semi-amplitude $K_2 = 18\pm4$\,km/s (see Sect.~\ref{subsubsec:helium_ionized}). 
The absorption strength also varies with the orbital phase, as illustrated in Fig.~\ref{fig:He_abs_profile_variation}. The absorption is stronger for phases 0.6 to 1.0 and takes its deepest value at $\phi=0.85$ when the stream impacts the AD outer radius, $r_t$. Hence, the absorption lines come from the outer edge of the optically thick AD and could be the effect of the deflection of the $L_1$ stream at the bright spot \citep{L1_deflection_2019ApJ...870..112G}. 
When the stream bounces off over the disc, it creates an extended vertical extension after the stream-disc impact that induces a veiling of the inner AD for phases in the range $\phi=0.7-0.9$ \citep{veiling_2001MNRAS.322..499K}. Depending on the deflection angle, a radial-velocity shift between 3 to 6 times the WD semi-amplitude is predicted. 
Furthermore, veiling by disc overflow would provide a natural explanation for the fainter maximum reported in the ellipsoidal variations at $\phi=0.6-0.9$ during the SAP \citep{TCrB_prediction_outburst_2023MNRAS.524.3146S} and illustrated in Fig.~\ref{fig:Blightcurve} for the $B$-band light curve.

\subsection{Summary}
\begin{figure}[t]
\centering
    \includegraphics[trim=1.9cm 5.5cm 1cm 5cm, clip,width = 0.5\textwidth]{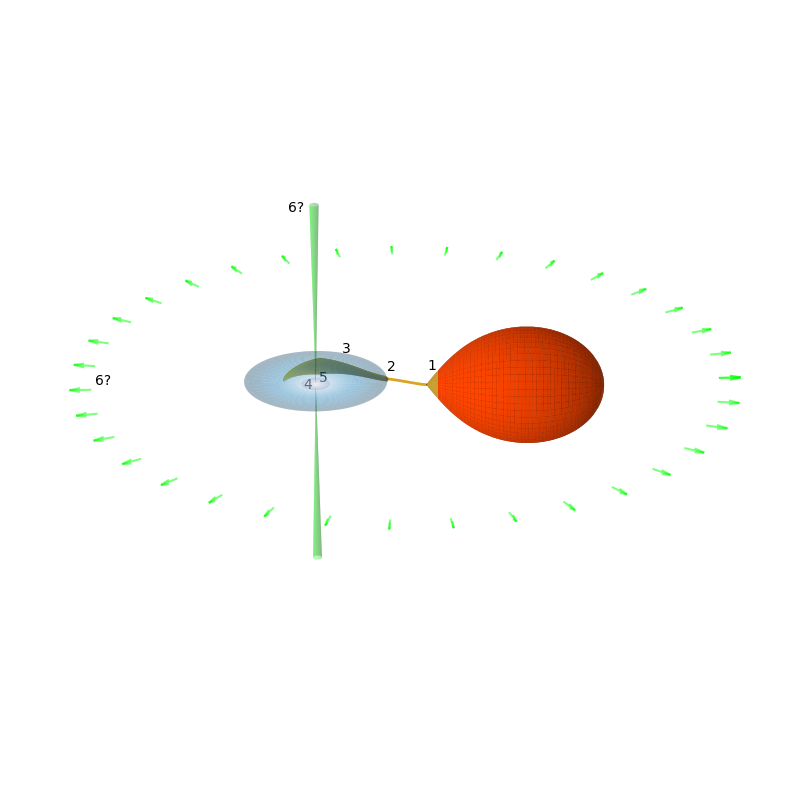}
    \caption{Schematic view of the system for $i$ = 65$^\circ$ and $q$ = 0.75 at $\phi=0.25$. The numbers displayed ($1 \rightarrow 6$) refer to Table ~\ref{tab:summary_table}.} 
    \label{fig:model}
\end{figure}

Different types of emission patterns are found: broad (\ion{H}{i} and \ion{He}{i} triplet) associated with the extended wind around the hot component, S-wave (\ion{O}{i} and \ion{He}{i} singlet) associated with the bright spot at the AD outer rim, single-peaked flickering emission (\ion{He}{ii}, \ion{N}{iii}) from the hot region near the boundary layer between the AD and the WD, and static double-peaked emission ([\ion{Ne}{iii}], [\ion{O}{iii}]) from the expanding nebula.
For the absorption lines, the absorption excess around $\phi=0.5$ (\ion{O}{i} triplet) is associated with an irradiated spot at the surface of the giant facing the WD. The stream overflowing an optically thick AD explains the absorption lines (\ion{He}{i} transitions) in phase with the hot component motion but with a larger velocity amplitude. The inferred locations of the different atomic transitions are summarized in Table \ref{tab:summary_table}, and a schematic view of the system is shown in Fig.~\ref{fig:model}.

\begin{table}[t]
    \centering
    \caption{Interacting sites in T~CrB system inferred from the spectroscopic monitoring during the SAP.}
    \begin{tabular}{rllrr}
    \hline
    \hline
        $n$ &Interaction site & lines &$T$ [kK]& $t_{p}$ [d]\\
        \hline      
        1&irradiation spot&\ion{O}{i} (abs)& [5, 10] & 1289\\
        & &$\lambda$7772$-$7775 & & \\
        2 &bright spot& \ion{He}{I} $\lambda$6678  &>20$^a$& 481\\
        & & \ion{O}{I} $\lambda$8446 & & \\
        3 &disc overflow & \ion{He}{I} (abs) & [9, 10]$^b$& $-$ \\ 
         & & ($\lambda$<4800) & &  \\
        4 &accretion wind& \ion{H}{i} $\lambda$ 6563&>20$^a$ &860\\
         & & \ion{He}{i} $\lambda$5578 & &\\
        5 &boundary layer & \ion{He}{ii} $\lambda$4686 &>60$^a$& 894\\
         & &  \ion{N}{iii} $\lambda$4641 & & \\
        6&nebula (jet or ring ?)& [\ion{O}{iii}], [\ion{Ne}{III}] & [10-50]& 210\\

         \hline
    \end{tabular}
    \tablebib{$a$: \cite{lithium_2020AJ....159..231W}, $b$: \cite{accretion_superactive_2023A&A...680L..18Z}}
    \tablefoot{$n$ refers to the different locations in Fig.~\ref{fig:model}. $t_{p}$ is the time of the peak maximum, taking $t=0$ as the start of the SAP (JD=2457023.5). $T$ provides a range or a lower bound on the temperature.}
    \label{tab:summary_table}
\end{table}

\section{Secular evolution}
\label{subsect:timing_lines}

\begin{figure}[t]
    \centering
    \includegraphics[width=\linewidth]{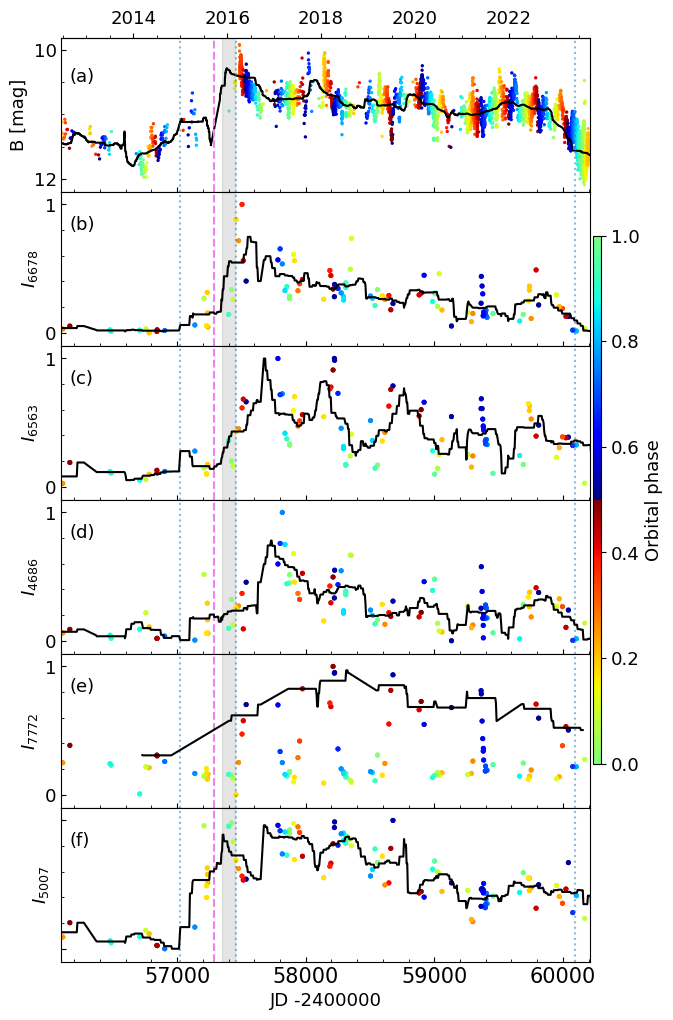}
    \caption{Relative intensity, $I_\lambda$, of the emission peak as a function of time for selected spectral lines, $\lambda$, and colored according to their orbital phase. The $I_\lambda$ values have been divided by their maximum value over the time series to set them on a [0,1]~interval, see text. The upper panel displays the $B$-band photometry.  For each panel, a 227-day rolling mean is drawn in black. The SAP's start, maximum, and end (vertical dotted lines) are shown. For the pink and gray lines, see the text.}
    \label{fig:timing}
\end{figure}

The previous section analyzed the orbital dependency of the spectral lines, probing the symbiotic activity to assess their location within the system. This section compares the photometric brightening with the long-term evolution of the spectral-lines intensity probing different interaction sites (see Table~\ref{tab:summary_table}). This is illustrated in Fig.~\ref{fig:timing}, where the temporal evolution of the peak relative intensity ($I_\lambda$, obtained after normalizing by its maximum value over the temporal evolution) is shown for five representative spectral lines $\lambda$, and compared with the light curve in the $B$ band (Fig.~\ref{fig:timing}a) that sets the start, maximum (taken from \cite{Dwarf_nova_2023ApJ...953L...7I}), and end of the optical brightening (vertical dotted lines). The complete spectral profile evolution of the individual spectral line during the SAP can be found in Fig.~\ref{fig:timing_appendix}. 
As reported by \cite{Luna_BL_2019ApJ...880...94L}, the ratio of soft- over hard- emissions started after the beginning of the SAP (vertical gray region in Fig.~\ref{fig:timing}) and about 60 days after the UV flux maximum (pink dashed line). 
The peak intensity of each spectral line is correlated with the SAP, as their intensity increased strongly during 2015-2023 and abruptly declined at the end of the SAP. Since all these lines (except [\ion{O}{iii}], see Sect. \ref{subsubsec:forbideen_transition}) experience a strong variability with the orbital motion, their evolution should only be compared after averaging over an orbital cycle (black curves in Fig.~\ref{fig:timing}).

The intensity strength of \ion{He}{i} at 6678\AA\, (Fig.~\ref{fig:timing}b), probing the bright spot, is directly correlated with the $B$-band light curve: a fast rise reaching its maximum in 2016, and then followed by a slow decline until the end of the SAP. This correlation can be understood assuming the $B$-band flux is dominated by the emission from the AD outer edge during the SAP. The H$\alpha$ and \ion{He}{ii} emission lines (Fig.~\ref{fig:timing}c-d), probing the AD wind at the outer and inner edge, have their maximum delayed by about one year after the light-curve maximum.

The intensity $I_{7772}$ displayed in Fig.~\ref{fig:timing}e represents the relative depth of the oxygen absorption line, and the rolling mean is computed only taking into account data in the phase interval [0.4, 0.6]. The depth is strengthened during the SAP: the phase-averaged absorption depth curve is symmetrical and takes its maximum value at the middle of the SAP, around 2018. Therefore, it seems time-delayed compared to the emission lines probing the disc. This finding is further discussed in Sect.~\ref{sec:discussion}.

\begin{figure}[t]
    \centering
    \includegraphics[width=\linewidth]{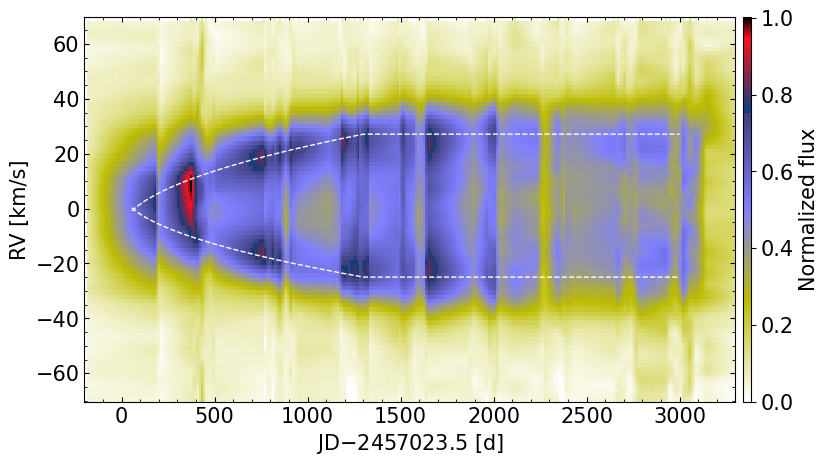}
    \caption{Temporal evolution of the [\ion{O}{iii}] profile. The SAP start is set at $t = 0$. The emission profiles are obtained after removing the giant contribution and interpolating between the observations. The evolution of the peak positions is represented by the dashed white line. The emergence and acceleration of the nebula is clearly seen. }
    \label{fig:Oiii_residual}
\end{figure}

In contrast, the emission of the nebular line [\ion{O}{iii}] (Fig.~\ref{fig:timing}f) attains its maximum values before the light curve maximum and even before the softening of the X-ray radiation when the boundary layer changes its optical thickness (vertical gray region in Fig.~\ref{fig:timing}).
In addition to its intensity variation,  the [\ion{O}{iii}] line underwent a profile variation during the first half of the SAP, illustrated in Fig.~\ref{fig:Oiii_residual}. The emission pattern was first single-peaked during the rise of the SAP, then changed to a double-peaked profile whose absolute radial velocity increased during the first 1000 days. The positions of the two peaks of the emission profile are fitted with two Lorentzian profiles and follow a $v \propto \sqrt{t}$ evolution during the first half of the SAP to reach their asymptotic value of (25,$-$23)~km/s implying that the ejecta continues its expansion with a frozen structure. The velocity stays constant until the end of the SAP when the line fades below the detection limit.
Assuming that the emission profile is associated with a localized amount of matter (e.g., blast wave or ionization front) launched and accelerated, the distance it traveled along the line-of-sight during the SAP (about 3000~days) would be about 40$\pm$10~au. Using the distance from \citealt{distance_bailer_jones_2021AJ....161..147B} (see Sect.~\ref{sec:introduction}), it corresponds to a sky-projected displacement of about $106\pm26$~mas if the ejection goes in the polar direction and of about $50\pm12$~mas if the ejection travels in the equatorial plane. No evidence of any structure was found in the radio-emission maps of \cite{SAP_radiorange_2019ApJ...884....8L} obtained on 19/12/2016 because the traveled distance by the ejection in its accelerating phase (corresponding to SAP~+~718~d in Fig. \ref{fig:Oiii_residual}) would have been below the spatial resolution of the radio-image ($\sim63~$mas).

To summarize, the symbiotic lines at the different interaction sites dramatically strengthened during the SAP. Still, their evolution does not appear to be synchronous: the emission from the expanding nebula emerged first, then followed by emission probing the AD outer edge and wind. A delayed effect is seen in the lines probing the irradiation at the giant surface. If significant, these delays between the spectral lines are the signature of causal relationships and/or different timescales governing their evolution. This is discussed in the next section.

\section{Discussion}
\label{sec:discussion}
From a comparison of the pre-eruption photometric archives, it was shown that the same photometric behavior was found for each of the two recorded nova explosions (\citealt{super_active_phase_2016NewA...47....7M}, \citealt{Luna_2020ApJ...902L..14L}, \citealt{TCrB_prediction_outburst_2023MNRAS.524.3146S}):  A ten-year-long SAP followed by a dip before the eruption. Hence, the SAP is expected to be essential in triggering the eruption by increasing the mass accretion rate onto the WD surface (see Sect.~\ref{sec:introduction}). 
One of the remaining questions concerns the physical origin of the sudden transition from quiescence to SAP. The above spectroscopic analysis demonstrates that the recent SAP correlates with the bright spot enhancement. Such an increase in the bright spot luminosity should originate from the disc through a variation at its outer radius or from the donor through a mass transfer variation \citep{DIM_review_2020AdSpR..66.1004H}. 

Below, we discuss the spectroscopic evidence and the respective timescales of two possible origins of mass accretion enhancement in light of the usual instability scenarios existing for CVs. From there, we propose a scenario to explain the SAP of T~CrB. We end up briefly mentioning the superoutburst hypothesis.

\subsection{Mass-transfer variation} 
\label{subsection:mass_transfer}

Mass-transfer variation through $L_1$ should be initiated by an instability at the giant's surface. Mass-transfer enhancement is usually used to explain the stunted outbursts in some nova-like systems \citep{stunted_outburst_1998AJ....115.2527H}. A cyclic mass transfer could be either intrinsic (e.g., solar-type cycle) or externally driven (e.g., irradiation by a tilted AD). 
The spectroscopic analysis of \ion{O}{i} confirms that the SAP induces a change at the facing side of the giant that can be interpreted as the effect of enhanced irradiation. The effect of irradiation is found to be delayed with respect to the optical rise (see Sect.~\ref{subsect:timing_lines}), suggesting it traces a retroactive effect of the increased disc luminosity rather than being the cause of it. 

Assuming the irradiation reaches the atmospheric layer below the photosphere, the time scale for the heat to be evacuated at the surface should be controlled by the thermal adjustment timescale that varies with $(\Delta R /R_*)\times t_{KH}$ \citep{accretion_book_2002apa..book.....F}, where $\Delta R$ is the atmospheric depth reached by the irradiation, and $t_{KH}$ is the Kelvin-Helmholtz timescale ($t_{KH}= GM_*^2/R_*L_*\sim200$~years for the giant, taking a stellar luminosity, $L_*$, of $620\pm 120 L_\odot$; \citealt{1998MNRAS.296...77B}). 

Contrary to CVs where the main-sequence donor imposes thermal adjustment timescales larger than $10^4$yr, in the case of T~CrB, an irradiation reaching a few percent of the photospheric radius should induce a stellar expansion needed to radiate the excess flux in only a couple of years. The mass transfer at $L_1$ being very sensitive to a change in $\Delta R$ ($\dot{M} \propto \exp{\Delta R/ H_p}$), such an irradiation could cause significant fluctuations of the mass transfer rate.

\subsection{Disc instability model (DIM)} 
\label{subsection:DIM}
\begin{figure}[t]
    \centering
    \includegraphics[width=\linewidth]{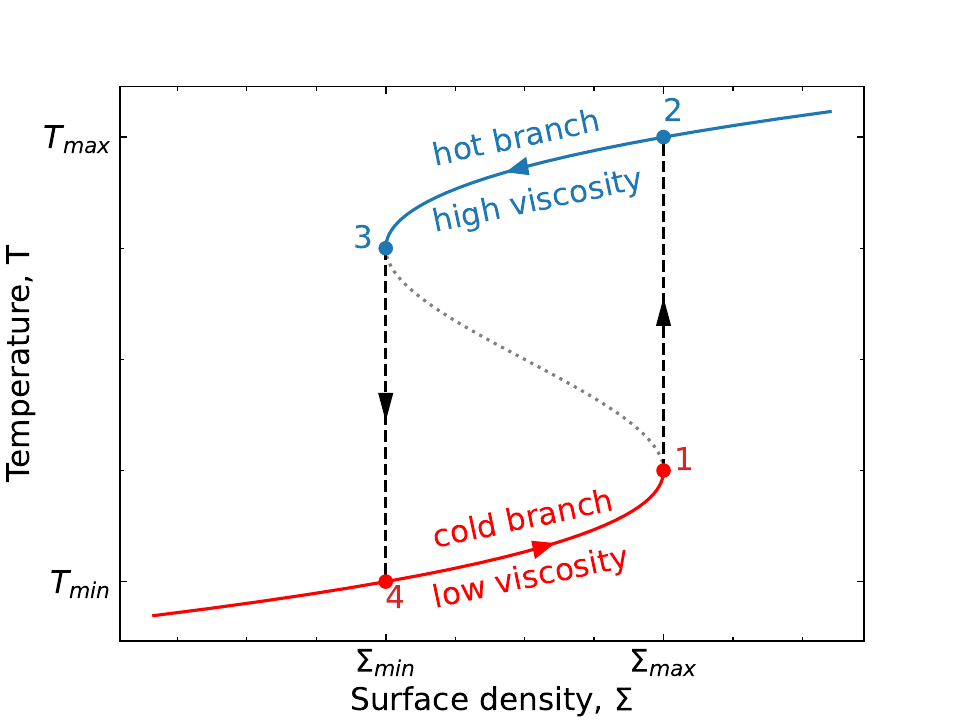}
    \caption{Schematic $\Sigma-T_{\rm eff}$ curve, showing the instability loop.}
    \label{fig:Scurve_sketch}
\end{figure}

DN outbursts are commonly explained by the thermal-viscous disc instability model (DIM, see the reviews of \citealt{DIM_1_2001NewAR..45..449L} and \citealt{DIM_review_2020AdSpR..66.1004H}) where the instability is attributed to partial hydrogen ionization within the disc that switches the disc from a low-viscosity cold state to a high-viscosity hot state. The canonical DIM and its cyclic behavior that we briefly recall below can be understood through the $\Sigma-T_{\rm eff}$ curve  (sketched in Fig.~\ref{fig:Scurve_sketch}), where $\Sigma$ is the disc surface density at a given radius. The two branches represent the thermally stable states that differ by their Shakura-Sunyaev viscosity parameter: $\alpha_h$ and $\alpha_c$ for the hot and cold branches, respectively. During the quiescent state, the AD lies on the cold branch of the curve, and its hydrogen is mainly neutral. With continuous mass transfer, matter accumulates in the disc and diffuses inward on a viscous timescale. The outburst begins when, at a given radius, the surface density reaches its maximal surface density $\Sigma_{max}$ (marked with 1 in Fig.~\ref{fig:Scurve_sketch}). The hydrogen ionization starts, and a heat wave propagates on a thermal timescale inward and outward ($1 \rightarrow 2$). During the outburst phase, the disc lies on the hot branch ($2 \rightarrow 3$), which is thermally stable, and enhanced mass accretion occurs at the WD surface. The hot branch is associated with a higher viscosity ($\alpha_h/\alpha_c$ is typically between 5$-$10) and, consequently, shorter viscous and thermal timescales than during the quiescence phase. The departure from the outburst phase happens when the surface density goes below $\Sigma_{min}$, and the AD returns to the cold branch, closing the instability loop ($3 \rightarrow 4$).

\cite{disc_instabilities_in_novae_2018MNRAS.481.5422B} have transposed the DIM used for DN outbursts to symbiotic systems. They conclude that the large ADs of symbiotic systems are still prone to thermal-viscous instability. Nevertheless, the larger dimension of the AD (T~CrB outer radius of $r_{t}\approx 5\times10^{12}$\,cm, see Sect. \ref{subsec:orbital_model}, compared to a mean outer radius of $5\times 10^{10}$\,cm for CVs, \citealt{DIM_review_2020AdSpR..66.1004H}) alters some of the properties of the disc instability
\footnote{For example, it should affect the propagation of heating and cooling waves during the outburst, preventing them from reaching the outer edge, while enhanced accretion rate at the boundary layer of the massive WD should induce a non-negligible irradiation effect on the disc outer edge. Also, large discs prevent outbursts from being initiated at the outer edge, implying that only ‘inside-out’ type outbursts (starting at the disc inner region and propagating outward) could occur \citep{disc_instabilities_in_novae_2018MNRAS.481.5422B}.} and its temporal evolution.
The timescales governing the disc equilibrium (dynamical, thermal, and viscous) scale with the disc geometry, roughly through the following sequence: $\alpha \, (H/R)^2  \, t_{vis}  \sim \alpha \,t_{th}  \sim t_{dyn} \sim r/v_K = \sqrt{ r^3/GM}$ \citep{accretion_book_2002apa..book.....F}.
Hence, a disc size larger by two orders of magnitude affects the different timescales by an increase of $\sim 1000 \times$. 
As a comparison,  DN experience outbursts lasting 1$-$10 days, between two and three orders of magnitude shorter than the SAP duration of $\sim3000$ days. 
The timescale for the rise of the outburst should be proportional to the disc's thermal timescale, $t_{th}$, while the time spent in the hot state is about one or two orders of magnitude larger as it scales with the viscous timescale (as $H/R$ < 1 for thin discs: \citealt{accretion_book_2002apa..book.....F}). 
So, the SAP shape (with a rise of $\sim$ 450 days and duration of $\sim3000$ days) is qualitatively compatible with the expected order of magnitude of $t_{th}$ and $t_{visc}$ during a DIM outburst extrapolated for the large AD of the T~CrB system, making the disc instability a viable candidate to explain the SAP properties. 
 
\subsection{The suggested scenario}
\label{subsection:candidate_scenario}
The AD properties inferred from the spectroscopic monitoring study agree with a thermally-unstable disc, as required for DIM. The delayed irradiation of the donor can be attributed to a (positive) retro-action feedback loop of the increased disc luminosity sustained by mass transfer from the irradiation. This supports the idea that a DN-type outburst (which probably started at the disc inner region) can account for the recent SAP of T~CrB and that the induced irradiation may have sustained the SAP by enhancing the mass transfer.

Further simulation of such DIM, considering the physical properties of T~CrB (mass transfer rate, AD extension, and SAP duration), should be foreseen to test the application of disc instability scenario for triggering the recurrent nova eruption and to asses how the stream-disc interaction and induced irradiation would affect the amplitude and duration of the outburst phase.  


\subsection{The SAP as a superoutburst ?} 
\label{subsection:superoutburst}

While the SAP has been recorded only twice in the T CrB system \citep{TCrB_prediction_outburst_2023MNRAS.524.3146S}, the system has shown several active states in the past (\citealt{IUE_data_1992ApJ...393..289S}, \citealt{RV_stanishev_2004A&A...415..609S}, \citealt{Dwarf_nova_2023ApJ...953L...7I}) of enhanced luminosity but reduced amplitude compared to the SAP (four times fainter in the $B$-band). 
This observation led \cite{Dwarf_nova_2023ApJ...953L...7I} to associate the recent SAP of T~CrB with superoutbursts as found in the SU~UMa-type DN. In their comparison, only these recurrent active phases of enhanced luminosity should be compared to the DN outbursts produced by thermal disc instability, whereas the SAP would instead be associated with a superoutburst. 
The superoutbursts occur in a subclass of CVs in addition to the regular outbursts but exhibit a longer duration and larger flux increase. They are suspected to be triggered by a tidal instability in an eccentric precessing AD \citep{osaki_Tidal_SU_UMa_1989PASJ...41.1005O}. However, such a mechanism cannot occur in the T~CrB system, as a tidally unstable disc requires a mass ratio smaller than 0.26 \citep{whitehurst_1988MNRAS.232...35W}. 
Suppose we want to explain the previously recorded active phases by the occurrence of DIM cycles. In that case, the increased strength and duration of the SAP with respect to active phases should be explained by a strong DIM reinforced by mass transfer, as proposed by \cite{SU_UMa_superoutburst_model_mass_transfer_2000A&A...353..244H} for SU~UMa-type CVs.

\section{Conclusions}
\label{sec:conclusions}
Thanks to a decade-long high-resolution spectral monitoring, we provide spectroscopic evidence of the accretion mechanism that governed the symbiotic system T CrB during its high state. This paper presents two main results.

In the first and main step, the orbital-phase dependency of the spectral lines probing the symbiotic activity in the optical range (mainly H, He, and O transitions) allowed us to investigate the accretion structure of the T  CrB system. To this end, Doppler tomography was performed for the first time on a symbiotic system. Different interaction sites have been resolved: the bright spot at the disc-stream impact, the stream-disc overflow, the AD wind, the irradiation spot at the red-giant surface, and the expanding nebula.  The mass transfer is dominated by Roche lobe overflow from the Roche-filling giant to an optically thick and viscously evolved AD surrounding the WD. 
These characteristics are reminiscent of what is spectroscopically observed in distinct classes of CVs, such as DN during their outburst phase and high-accreting nova-like systems, suggesting that the accretion mechanisms found in those short-period interacting binaries ($P<6$h for most CVs) can be transposed to symbiotic semi-detached binaries, despite their different donor type and their larger orbital period that imposes larger AD geometry and lower velocity regime.

In a second step, the time delay of the spectral lines emergence between 2015 and 2023 is investigated to understand their impact on the recent optical brightening event, referred to here as the super-active phase, SAP. We found that the rise of the optical flux is concurrent with the radial acceleration of the forbidden transitions probing the nebula. We interpreted this as a signature of bipolar ejection launched at the inner disc edge (e.g., blast wave at the WD boundary layer) whose interaction with the circumstellar medium leads to the excitation of forbidden lines that underwent heating and acceleration. Additionally, the evolution of the spectral line intensity suggests that the SAP of the T~CrB system was triggered in the inner part of the disc, increasing the disc temperature and the mass accretion rate at the WD surface, as previously suggested by \cite{Luna_BL_2019ApJ...880...94L} with X-rays observations. The higher temperature in the disc later affected the donor properties by inducing time-delayed irradiation at its surface. The emergence of the irradiation spot should have further enhanced the mass transfer and sustained the SAP. Hence, we suggest that the irradiation-induced mass-transfer enhancement has altered the duration or amplitude of the SAP phase, which could have been initiated by disc instability, as found in DN outbursts. The spectral signatures of the SAP abruptly stop in 2023, implying that the AD has return to a quiet phase of reduced activity and lower temperature.
This work consistently depicts the system's initial conditions before its nova eruption. With its imminent eruption, multi-wavelength monitoring (from $\gamma$-rays to radio range) will provide a unique view of its secular dynamics and the fate of the AD.


\begin{acknowledgements}
       We thank Jan Kàra for providing us with useful insights on Doppler maps.  
       L.P. is a FNRS research fellow. 
A.J. acknowledges support from the {\it Fonds de la Recherche Fondamentale Collective} (FNRS, F.R.F.C.) of Belgium
through grant PDR T.0115.23, and from FNRS-F.R.S. through grant CDR J.0100.21 ('ASTRO-HERMES'). 
L.P. and A.J. are members of BLU-ULB, the interfaculty research group focusing on space research at ULB - Université libre de Bruxelles.
Based on observations obtained with the HERMES spectrograph, which is supported by the Research Foundation - Flanders (FWO), Belgium, the Research Council of KU Leuven, Belgium, the Fonds National de la Recherche Scientifique (F.R.S.- FNRS), Belgium, the Royal Observatory of Belgium, the Observatoire de Genève, Switzerland and the Thüringer Landessternwarte Tautenburg, Germany. We acknowledge all observers of the HERMES consortium who contributed to obtaining the time series. 
We acknowledge with thanks the variable star observations from the AAVSO International Database contributed by observers worldwide and used in this research.
Use was made of the Simbad database, operated at the CDS, Strasbourg, France, and of NASA's Astrophysics Data System Bibliographic Services. This research made use of Numpy, Matplotlib, SciPy, scikit-image, and Astropy, a community-developed core Python package for Astronomy \citep{Astropy_2018AJ....156..123A}. 

\end{acknowledgements}

%
%

\bibliographystyle{aa} 
\bibliography{references} 

\FloatBarrier

\begin{appendix} 
\section{Photometry}
\begin{figure}[ht]
    \centering
    \includegraphics[width=\linewidth]{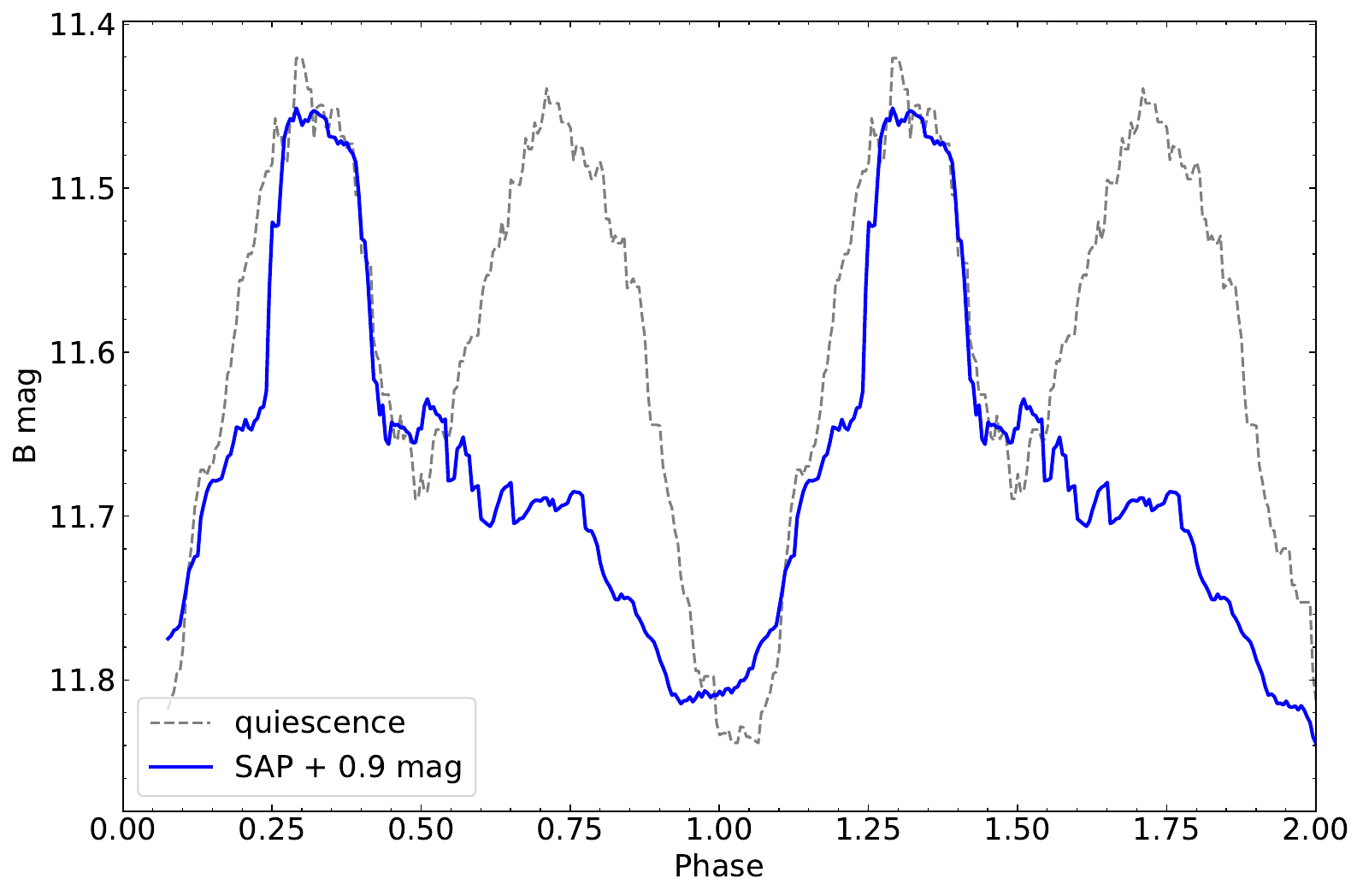}
    \caption{Phase-folded $B$-band light curve. The observed phases are displayed twice to guide the eyes.}
    \label{fig:Blightcurve}
\end{figure}

Figure \ref{fig:Blightcurve} represents the $B$-band photometry from AAVSO phase-folded and phase-averaged using a rolling mean of window $\Delta \phi= 0.15$ and step of $\Delta\phi=0.005$. The data of the SAP covers the JD-range 2457157$-$2459945, and the quiescence covers the JD-range 2450634$-$2456779.
The SAP is associated with a brightening of about 0.9~mag in the blue. 

Note how the second maximum (at phase 0.6$-$0.9) is obliterated during the SAP compared to the quiescent phase. This is caused by absorption from matter flowing over the disc (see Sect.~\ref{subsubsec:from_the_accretor} and flow labeled "3" in Fig.~\ref{fig:model}).
\section{Observation log}
\label{appendix:observation_log}
 Table \ref{tab:RV_data} lists the radial velocities from the HERMES spectrograph used in Sect.~\ref{sec:system_properties}. 
\begin{table}[h]
\caption{Radial velocities of T~CrB measured with HERMES (barycentric correction applied).}
    \centering
    \begin{tabular}{rrr|rrr}
    \hline
    \hline
        HJD & RV &$O-C$ &HJD & RV &$O-C$ \\
        2400000+ & [km/s] & [km/s] & 2400000+ & [km/s] & [km/s]\\
    \hline
55575.74 & -43.65 &  -1.19 &58355.39 & -11.68 &  0.57 \\
56110.39 & -3.60 &  0.98 &58503.78 & -52.74 &  -0.40 \\
56165.42 & -26.49 &  0.96 &58518.73 & -49.76 &  -0.40 \\
56478.41 & -46.50 &  -0.21 &58538.66 & -39.68 &  0.26 \\
56486.40 & -42.56 &  -0.30 &58558.60 & -26.87 &  0.13 \\
56756.70 & -15.88 &  -0.66 &58595.52 & -7.74 &  -0.52 \\
56782.48 & -6.07 &  -0.66 &58615.49 & -4.40 &  0.24 \\
56842.42 & -24.69 &  -0.92 &58646.54 & -14.22 &  -0.13 \\
56844.42 & -26.20 &  -1.15 &58661.42 & -22.97 &  -0.22 \\
56903.35 & -53.00 &  -0.68 &58678.44 & -34.00 &  -0.30 \\
57137.50 & -52.45 &  -0.06 &58881.70 & -18.63 &  -0.33 \\
57208.42 & -18.02 &  -0.83 &58897.71 & -28.18 &  0.17 \\
57228.43 & -7.96 &  -0.23 &58920.61 & -42.20 &  0.13 \\
57229.48 & -7.47 &  -0.07 &58986.54 & -44.71 &  -0.75 \\
57234.41 & -6.53 &  -0.47 &59002.49 & -35.34 &  -0.82 \\
57236.41 & -6.43 &  -0.80 &59025.41 & -20.04 &  -0.65 \\
57239.39 & -5.86 &  -0.73 &59037.40 & -13.11 &  -0.50 \\
57400.79 & -39.51 &  0.43 &59059.41 & -5.43 &  -0.21 \\
57422.69 & -25.42 &  0.25 &59133.33 & -33.61 &  -0.08 \\
57435.69 & -16.55 &  0.83 &59259.73 & -14.90 &  0.50 \\
57477.59 & -3.55 &  1.10 &59288.64 & -4.40 &  0.59 \\
57504.53 & -11.22 &  0.82 &59299.62 & -4.60 &  0.14 \\
57514.45 & -16.24 &  1.03 &59363.52 & -34.88 &  0.29 \\
57536.53 & -30.51 &  0.61 &59365.50 & -36.10 &  0.30 \\
57783.73 & -43.28 &  -0.42 &59371.39 & -39.57 &  0.34 \\
57800.72 & -49.57 &  0.36 &59376.48 & -42.53 &  0.18 \\
57817.74 & -51.79 &  0.67 &59380.48 & -44.39 &  0.34 \\
57838.57 & -48.46 &  0.00 &59381.46 & -44.76 &  0.43 \\
57852.62 & -41.63 &  0.22 &59383.41 & -45.90 &  0.19 \\
57874.65 & -27.75 &  0.06 &59396.43 & -50.78 &  -0.12 \\
57874.66 & -27.78 &  0.03 &59398.39 & -51.17 &  -0.05 \\
57905.44 & -9.63 &  0.26 &59401.39 & -51.47 &  0.23 \\
57912.47 & -7.17 &  0.15 &59404.40 & -52.01 &  0.10 \\
57939.44 & -5.61 &  -0.23 &59410.40 & -52.52 &  -0.06 \\
57952.42 & -8.81 &  0.14 &59456.38 & -34.48 &  0.85 \\
57974.40 & -20.34 &  -0.25 &59663.65 & -46.10 &  0.55 \\
58139.76 & -8.40 &  -0.99 &59691.73 & -29.88 &  0.31 \\
58187.61 & -12.89 &  -0.68 &59731.55 & -6.94 &  0.87 \\
58195.58 & -16.17 &  0.19 &59739.45 & -5.24 &  0.46 \\
58212.67 & -26.87 &  -0.02 &59740.43 & -4.89 &  0.61 \\
58222.65 & -33.32 &  -0.03 &59753.42 & -4.14 &  0.51 \\
58224.49 & -34.61 &  -0.15 &59791.39 & -18.03 &  -0.10 \\
58250.60 & -48.34 &  -0.17 &59811.41 & -30.70 &  -0.15 \\
58275.59 & -52.41 &  -0.03 &59949.79 & -11.90 &  -0.30 \\
58293.43 & -48.50 &  0.07 &59971.74 & -5.53 &  -0.58 \\
58294.45 & -48.16 &  0.03 &59996.79 & -7.97 &  -0.34 \\
58311.41 & -39.56 &  0.18 &60023.67 & -21.17 &  -0.41 \\
58311.42 & -39.12 &  0.61 &60042.52 & -33.29 &  -0.46 \\
58311.44 & -39.55 &  0.18 &60076.48 & -50.39 &  -0.46 \\
58346.44 & -16.54 &  0.57 &60104.56 & -51.22 &  0.07 \\
    \hline
   \end{tabular}
    \label{tab:RV_data}
\end{table}

\FloatBarrier

\section{Spectra and Doppler maps}
\label{appendix:spetra_and_Doppler_maps}
\begin{figure*}
    \includegraphics[width = 0.3\textwidth]{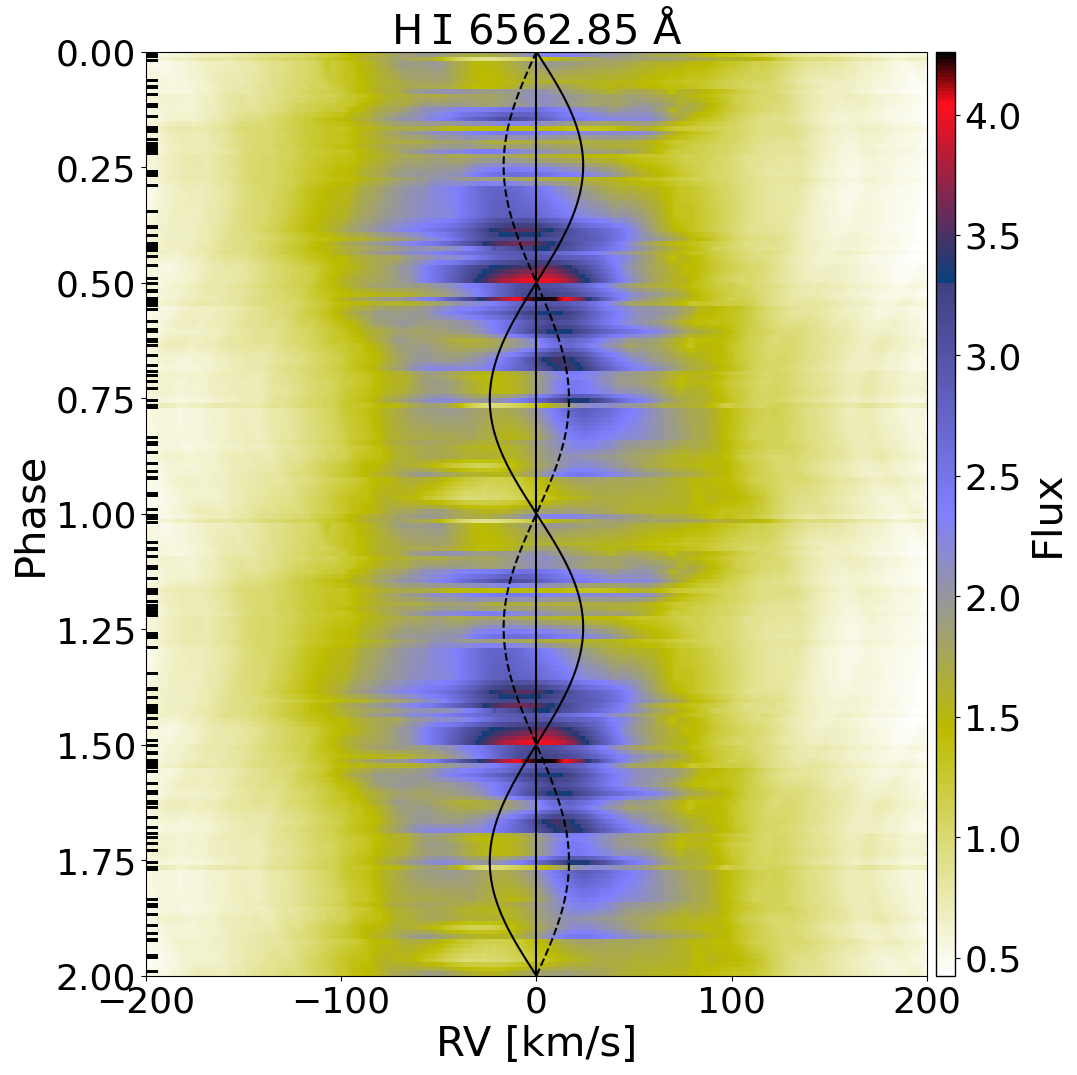}\hfill
    \includegraphics[width = 0.29\textwidth]{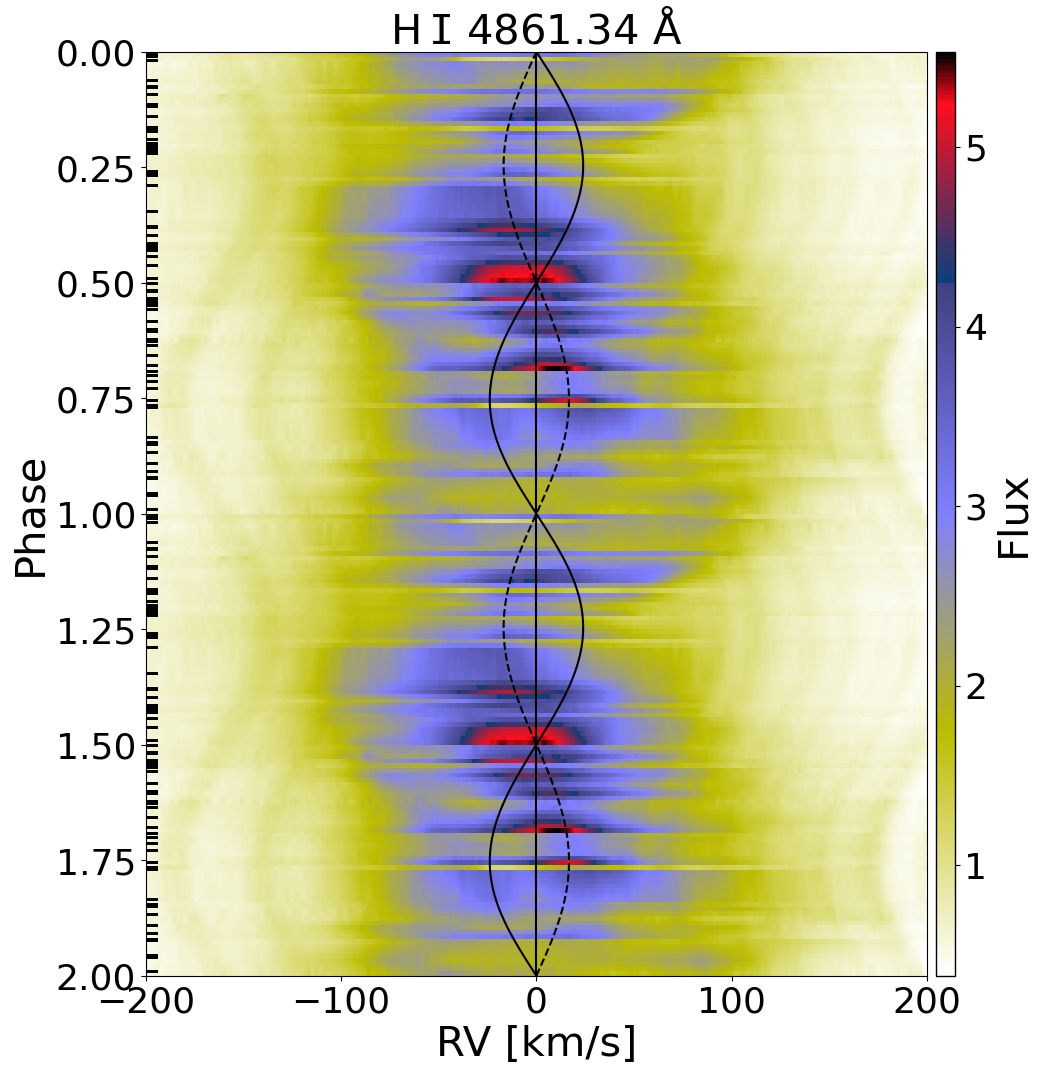}\hfill
    \includegraphics[width = 0.3\textwidth]{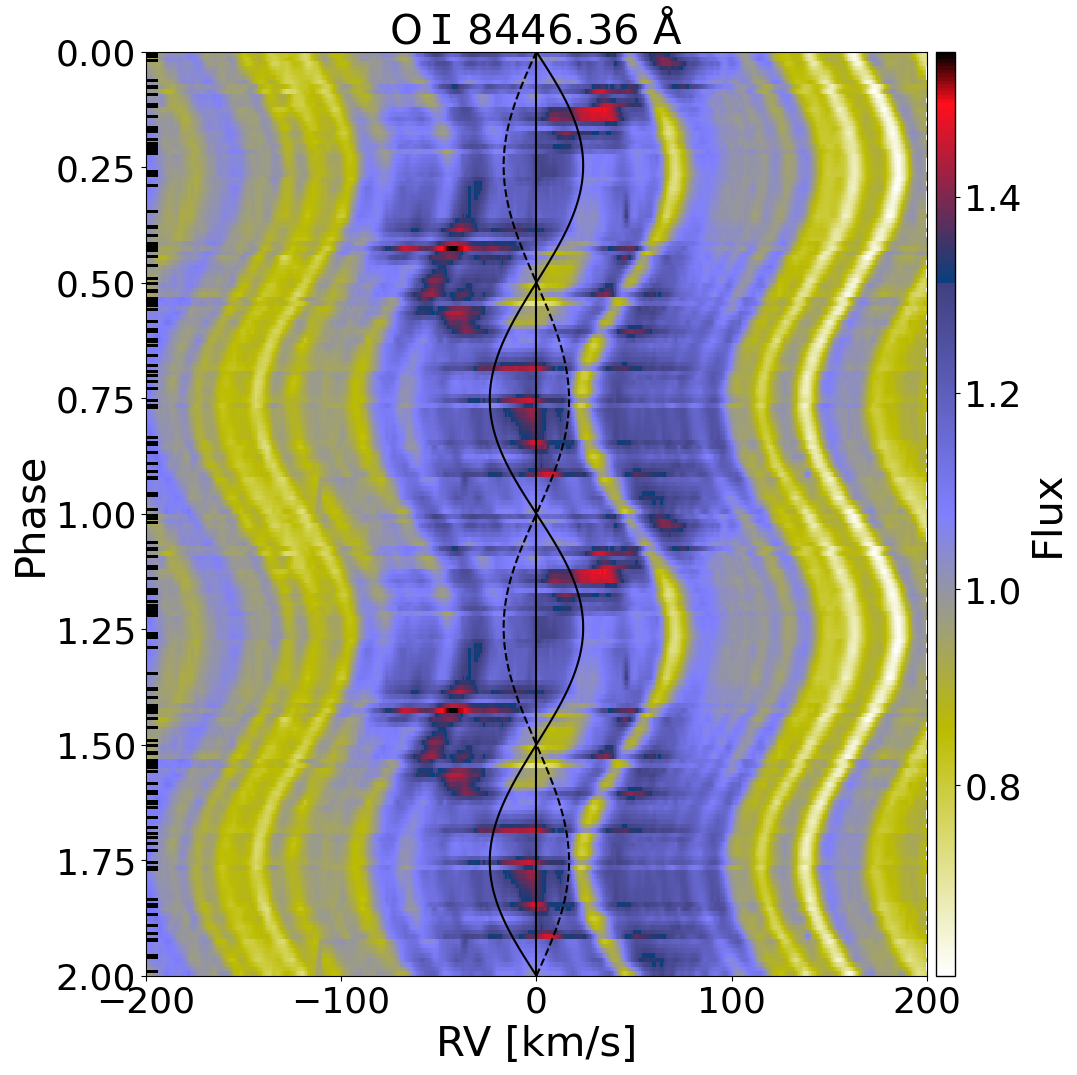}\hfill

    \includegraphics[width = 0.3\textwidth]{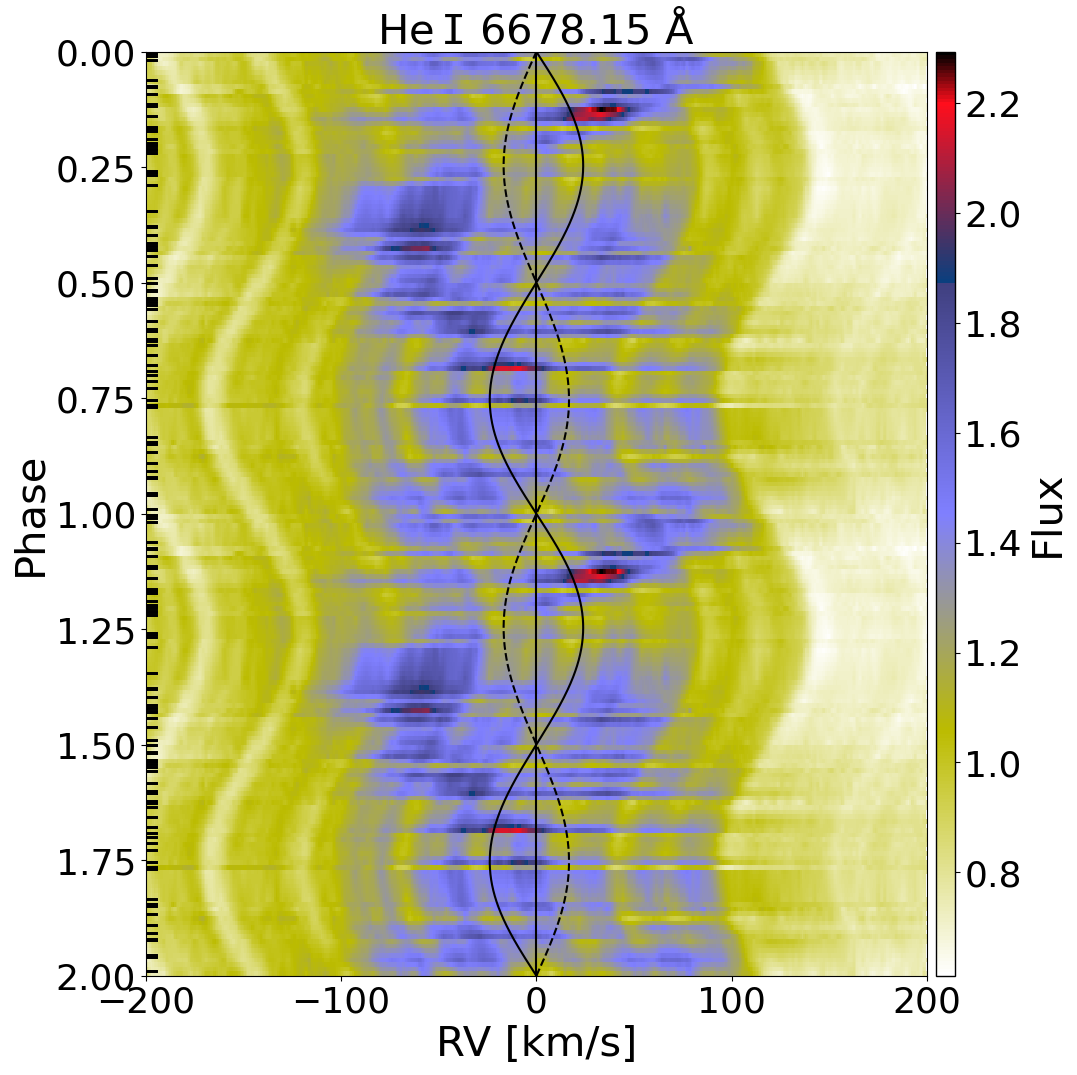}\hfill
    \includegraphics[width = 0.3\textwidth]{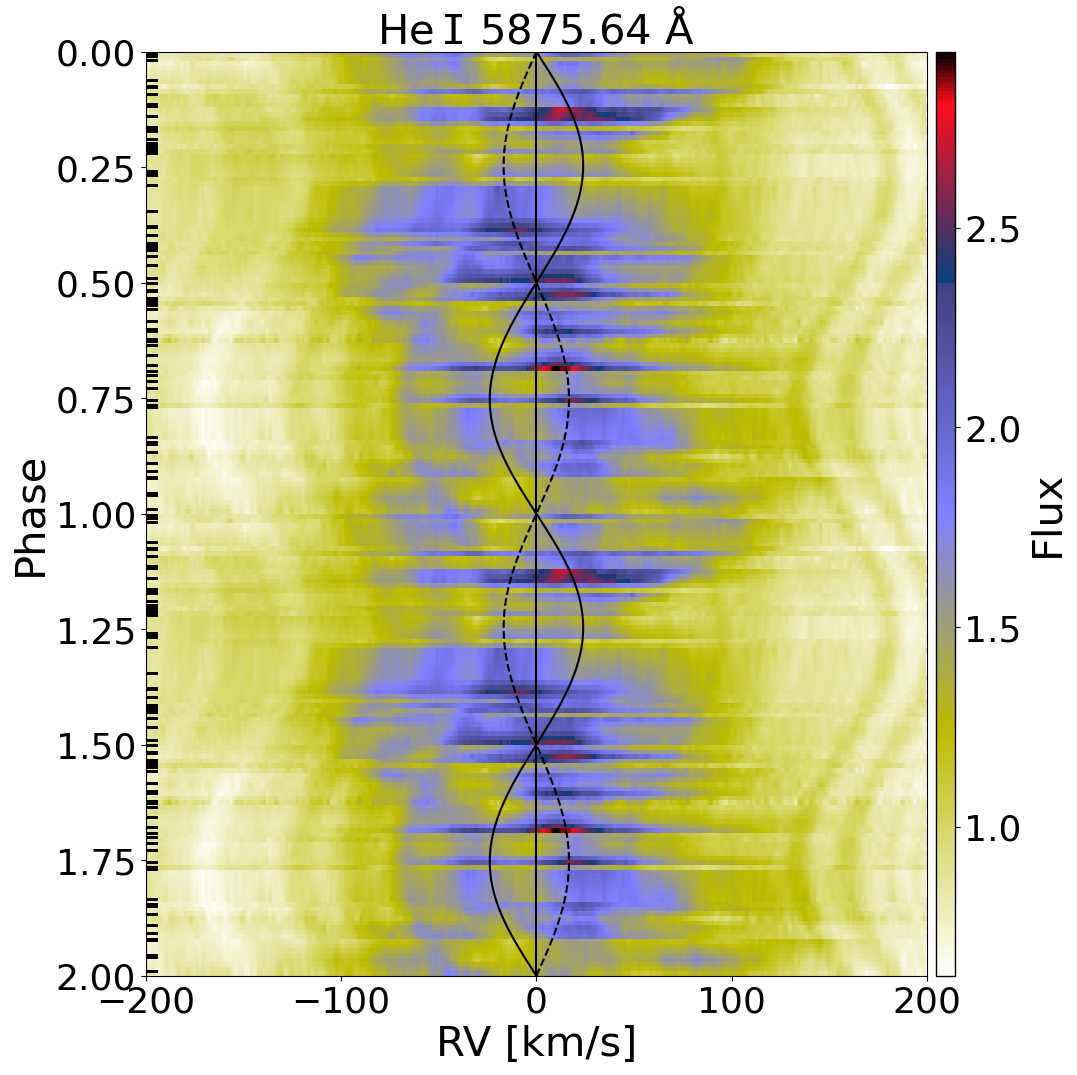}\hfill
    \includegraphics[width = 0.3\textwidth]{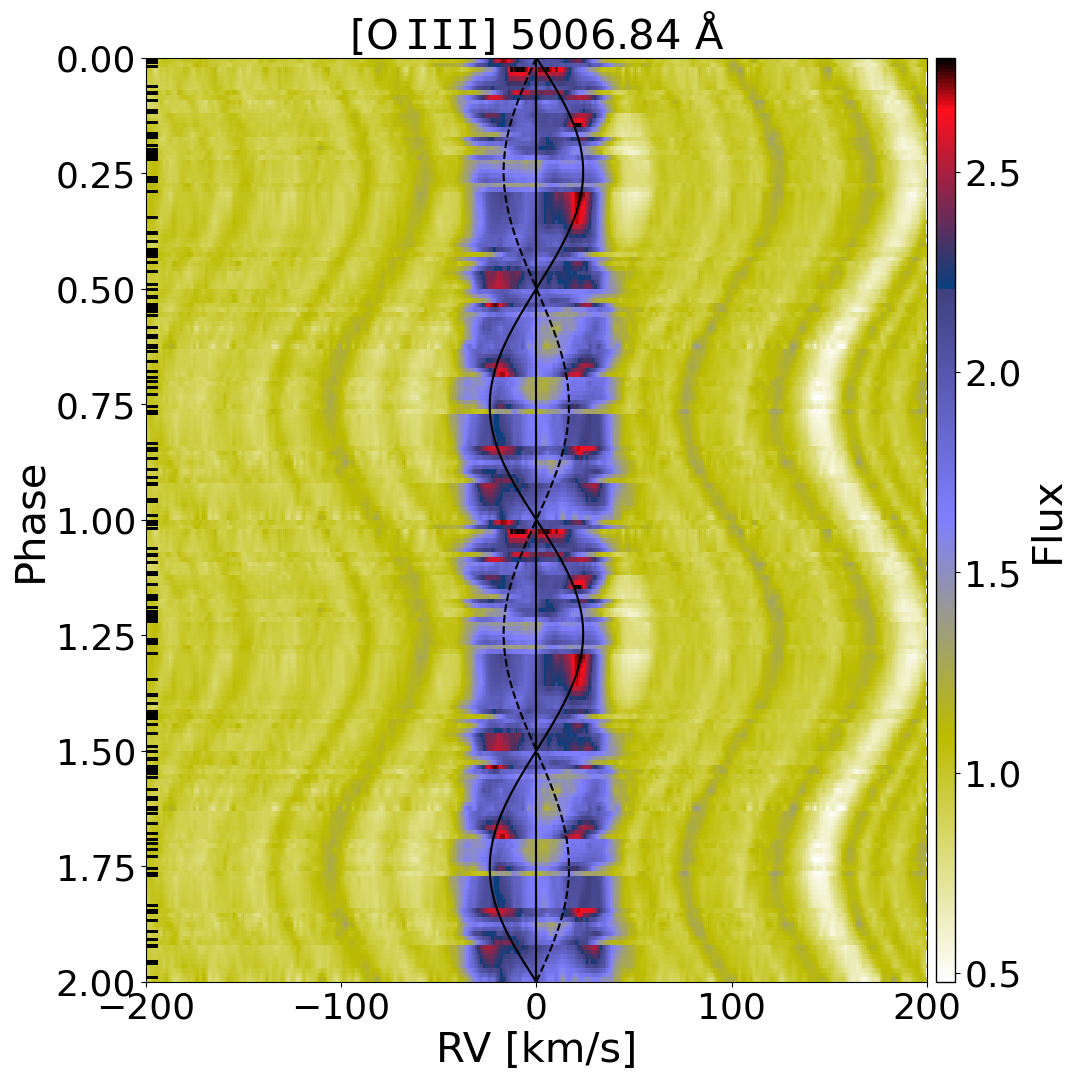}\hfill
    
    \includegraphics[width = 0.3\textwidth]{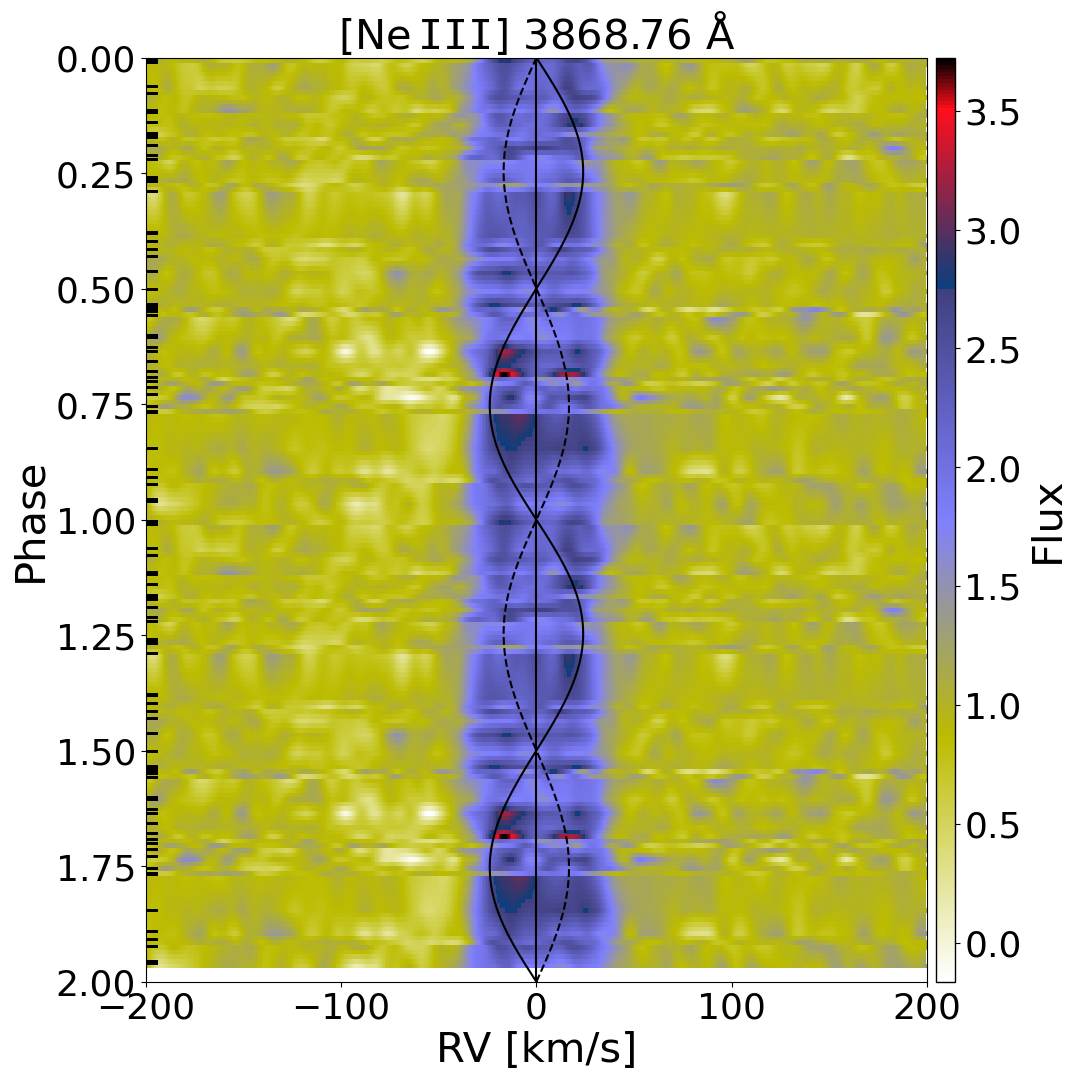}\hfill
    \includegraphics[width = 0.3\textwidth]{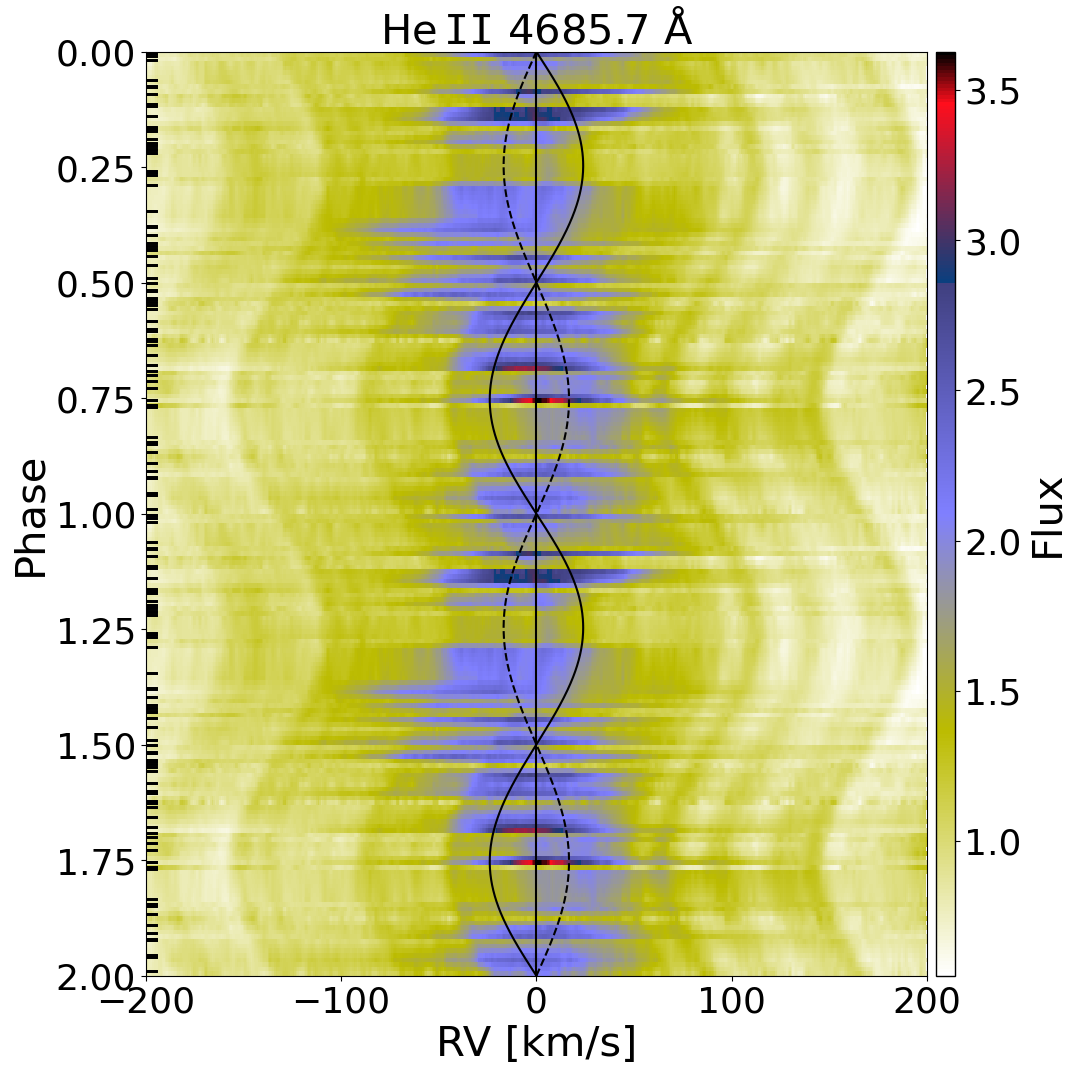}\hfill
    \includegraphics[width = 0.3\textwidth]{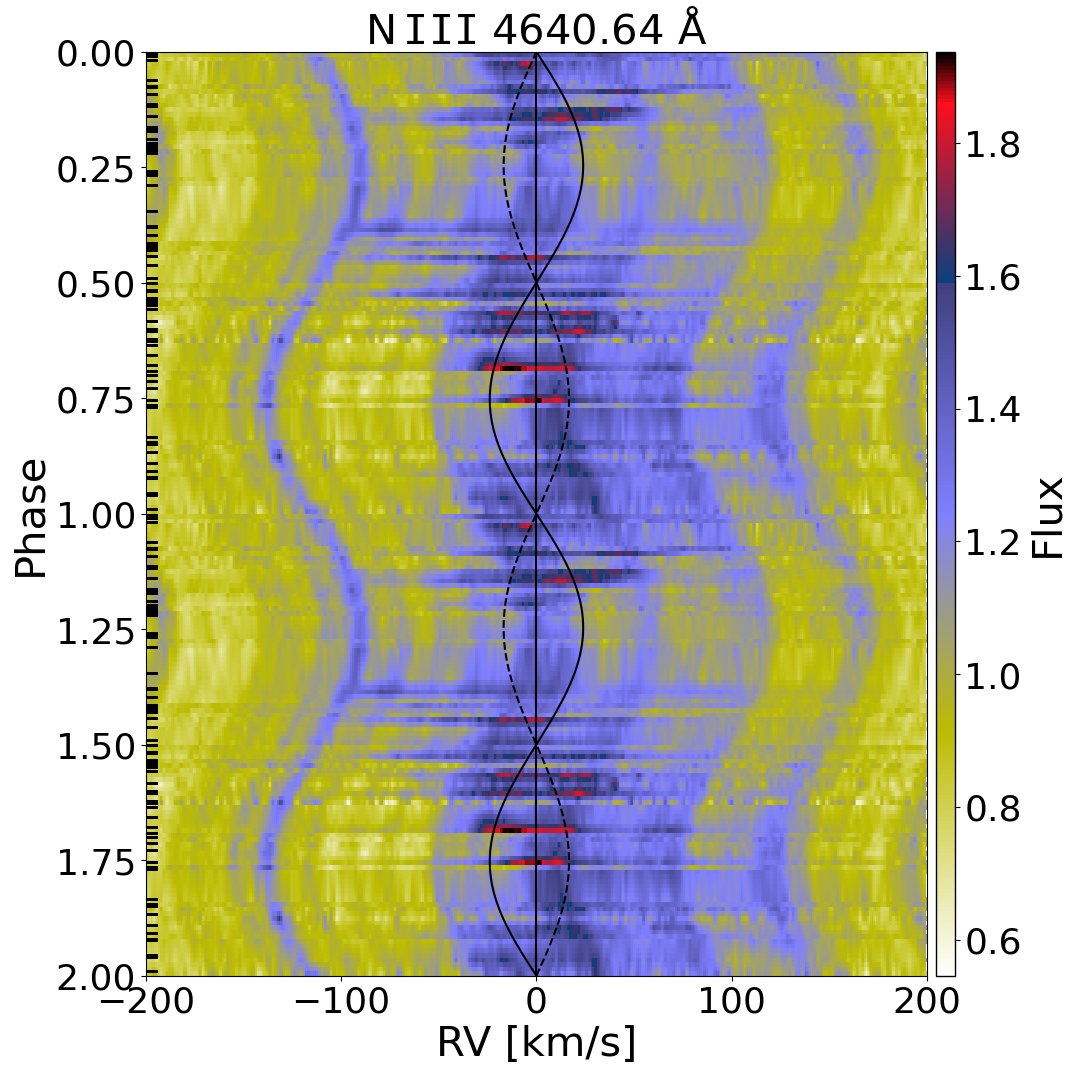}\hfill

    \includegraphics[width = 0.3\textwidth]{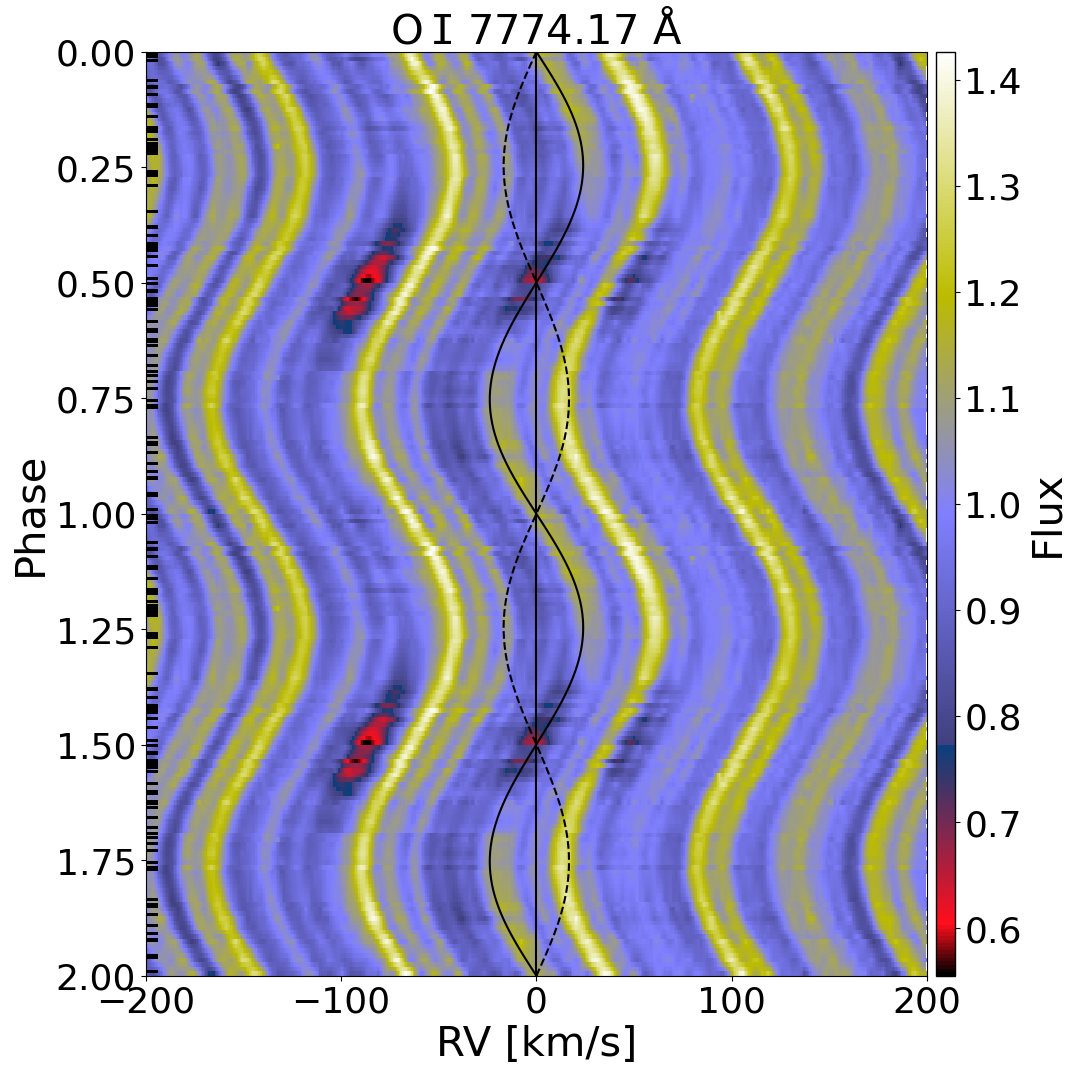}\hfill
    \includegraphics[width = 0.3\textwidth]{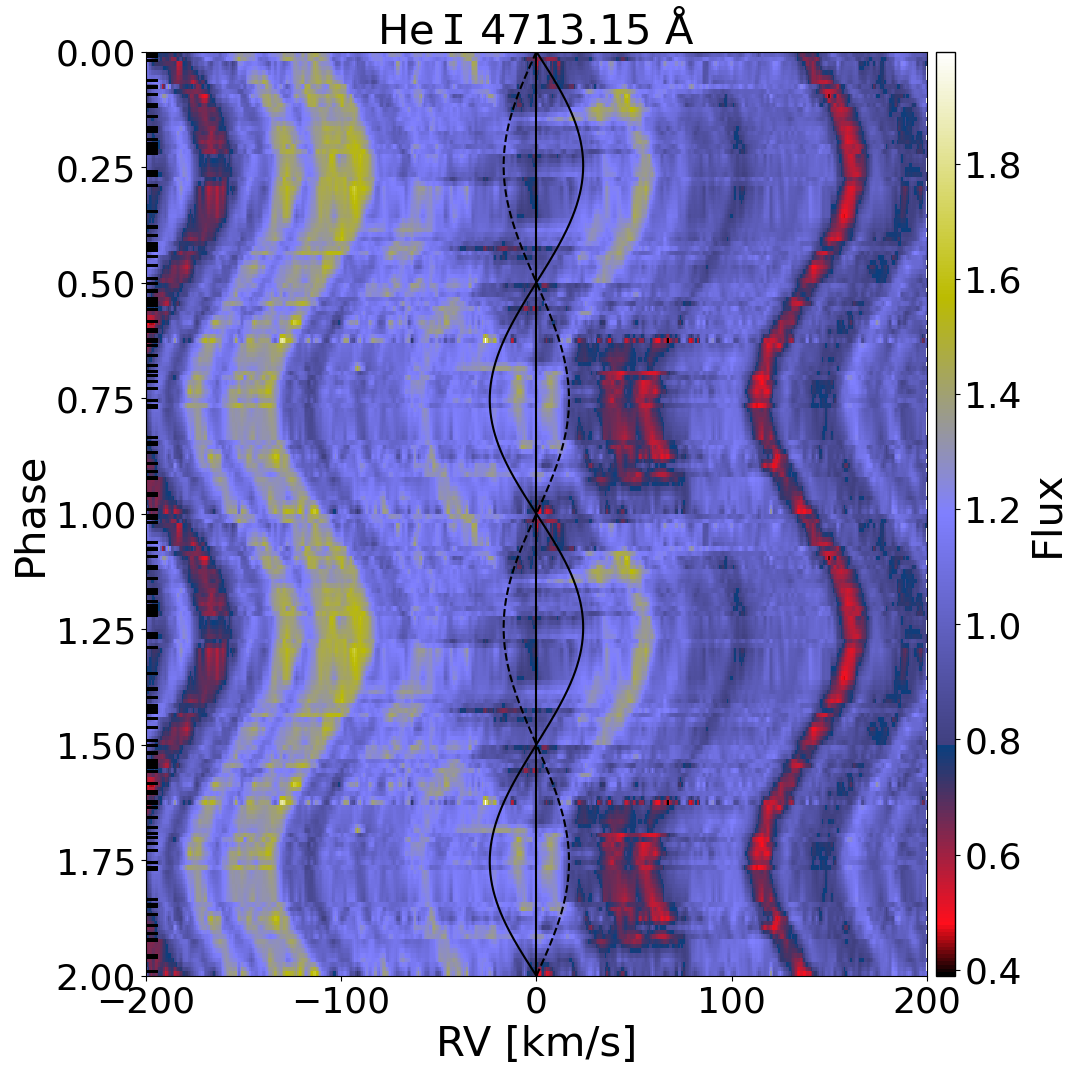}\hfill 
    \includegraphics[width = 0.3\textwidth]{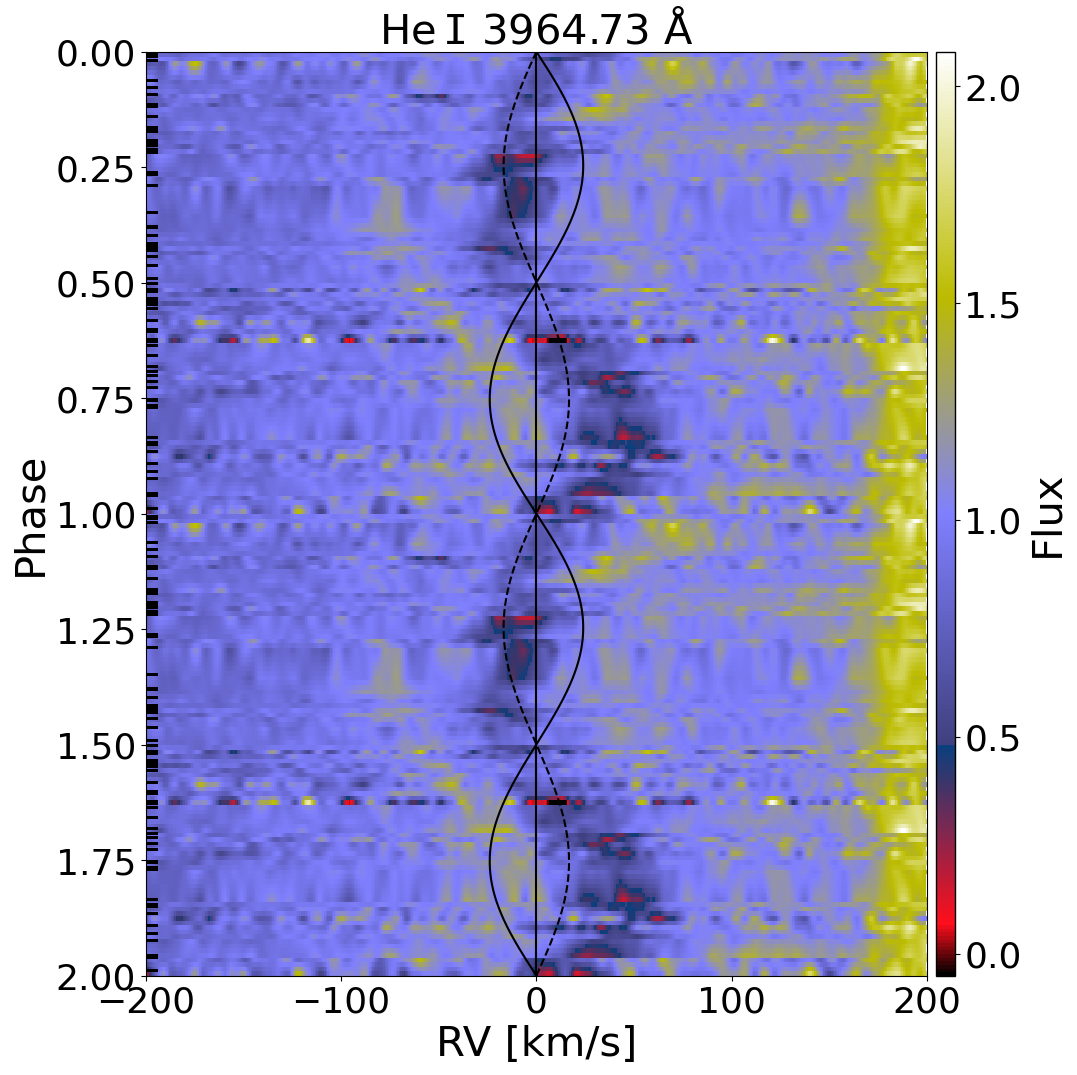}\hfill

    \caption{Dynamic spectra of T~CrB as a function of the orbital phase during the SAP. Each panel shows the observed phases twice to guide the eyes, with time running downward. The color represents the pseudo-continuum-normalized fluxes taken as the median value of the spectral window. The black curve represents the primary motion, while the vertical line marks the systemic velocity, and the horizontal lines on the left indicate the observed phases.}
    \label{fig:spectra_interpol}
\end{figure*}

\begin{figure*}
    \includegraphics[width = 0.3\textwidth]{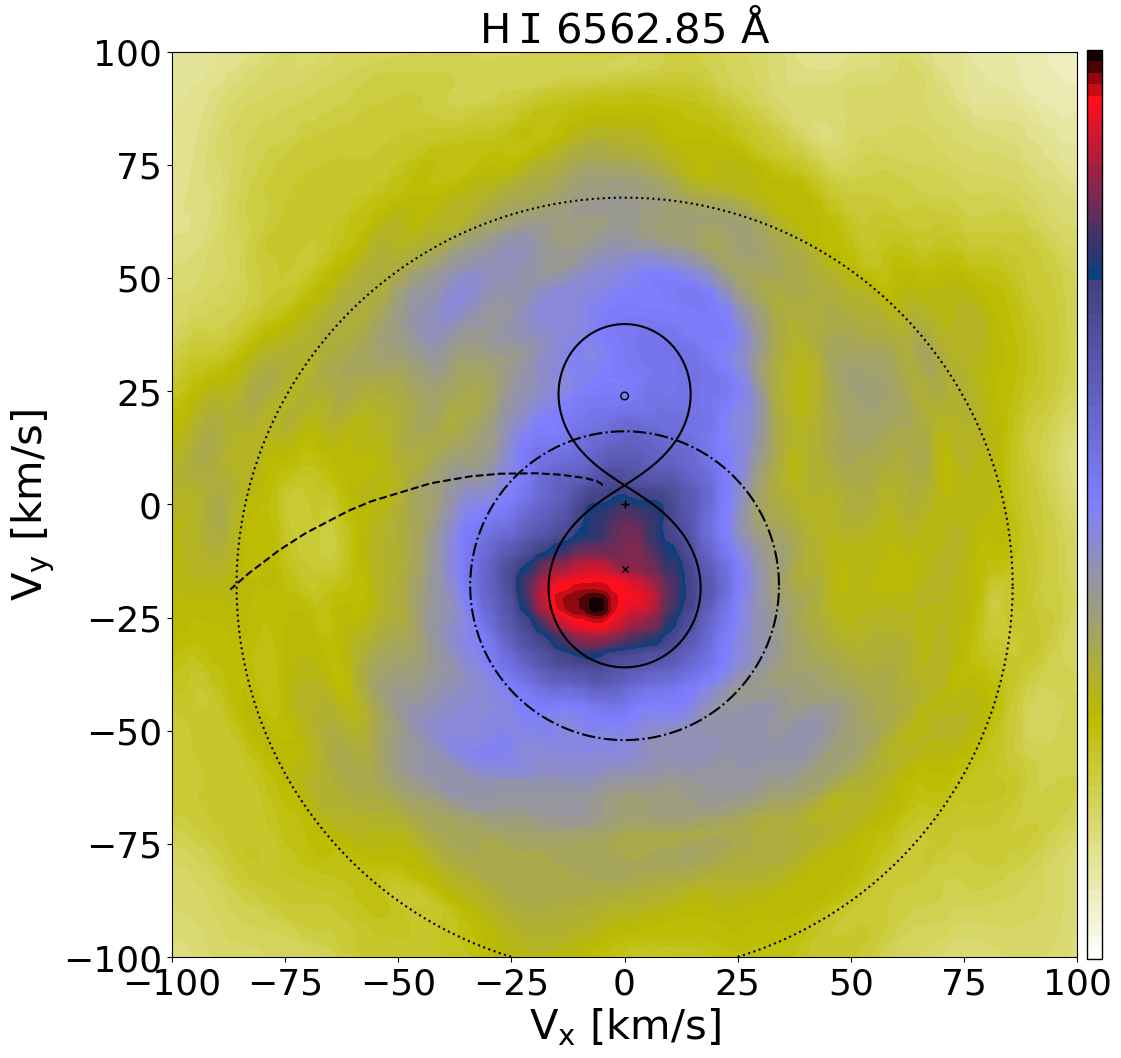}\hfill
    \includegraphics[width = 0.3\textwidth]{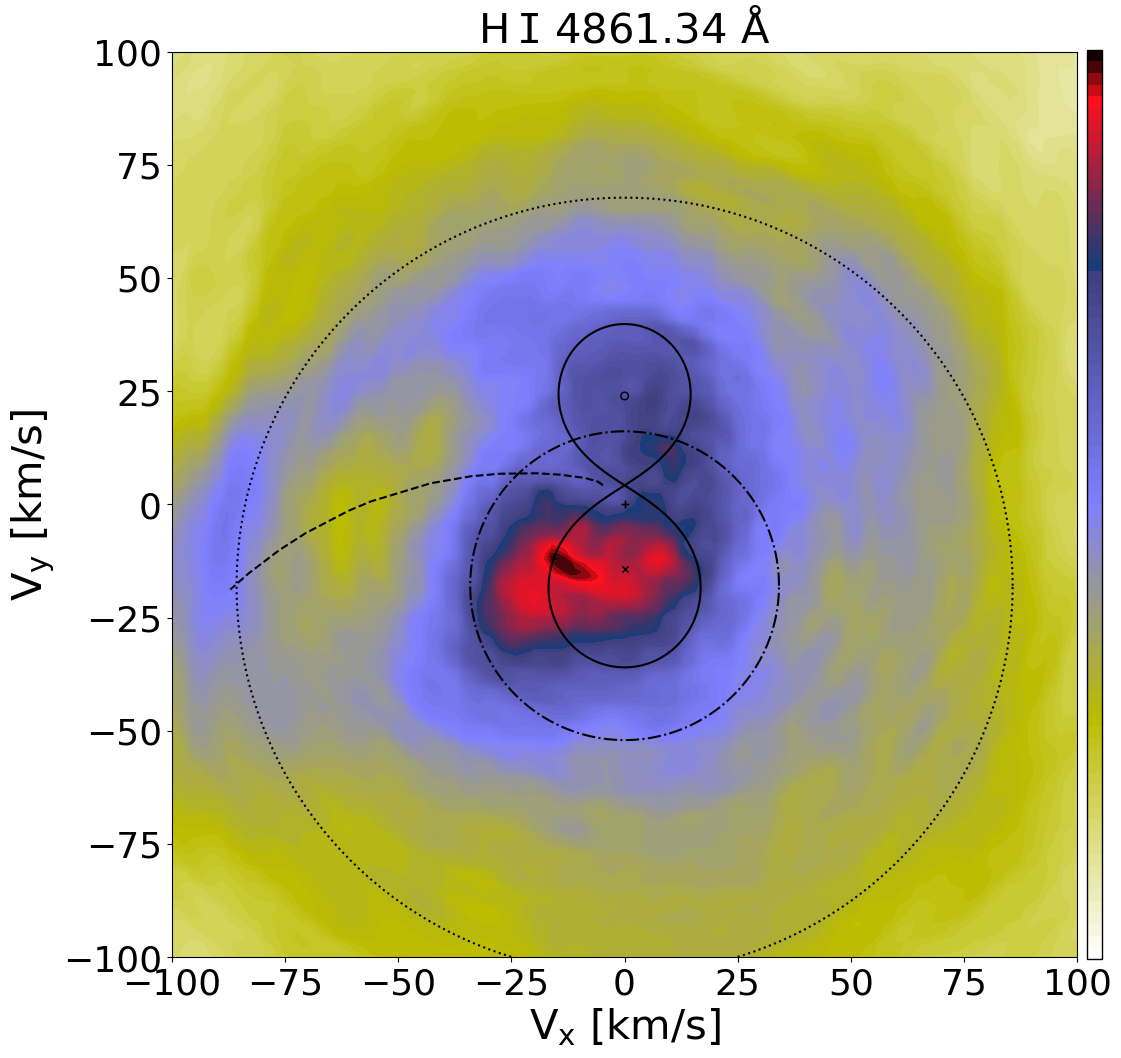}\hfill
     \includegraphics[width = 0.3\textwidth]{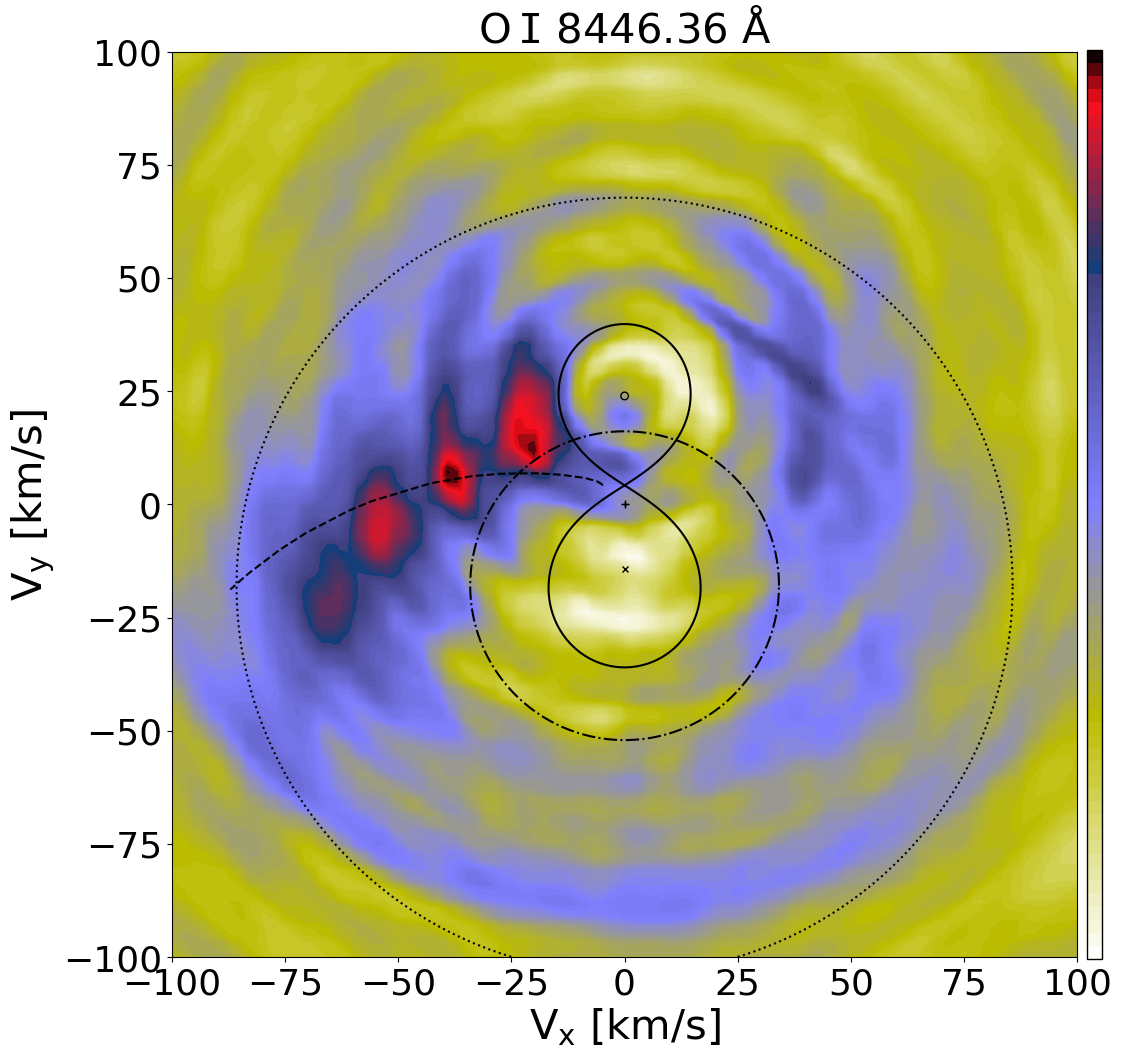}\hfill

    \includegraphics[width = 0.3\textwidth]{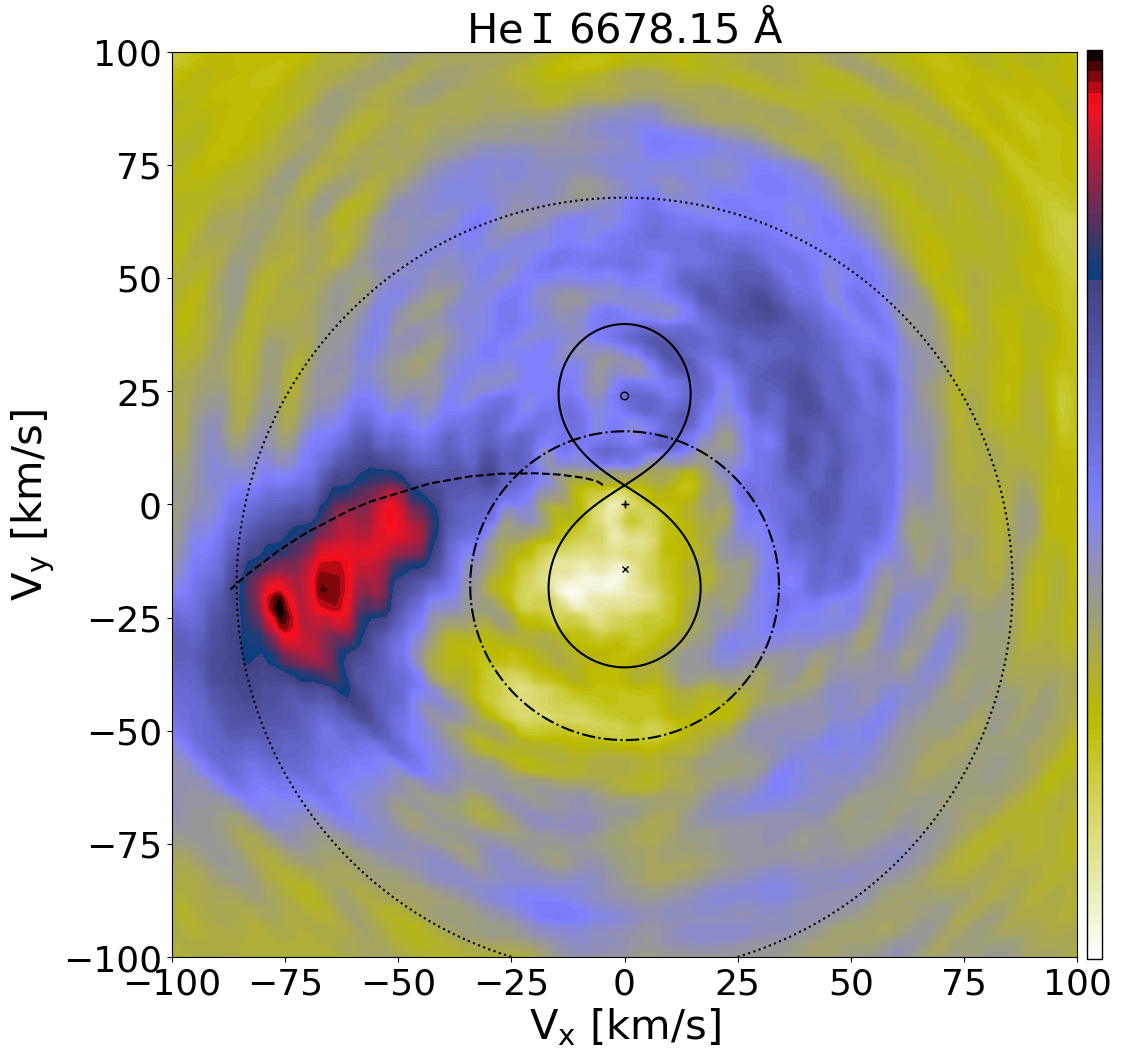}\hfill
    \includegraphics[width = 0.3\textwidth]{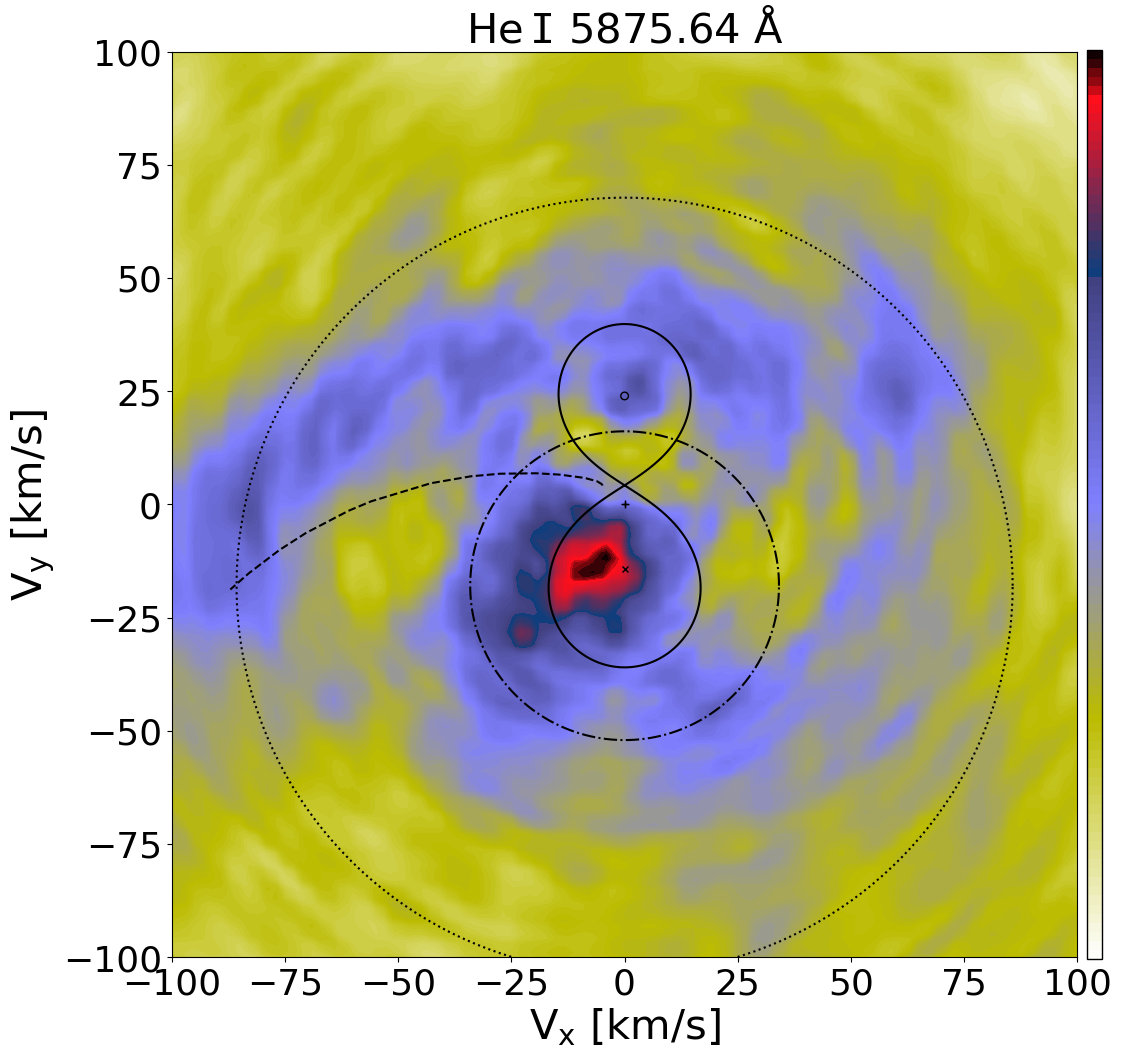}\hfill 
    \includegraphics[width = 0.3\textwidth]{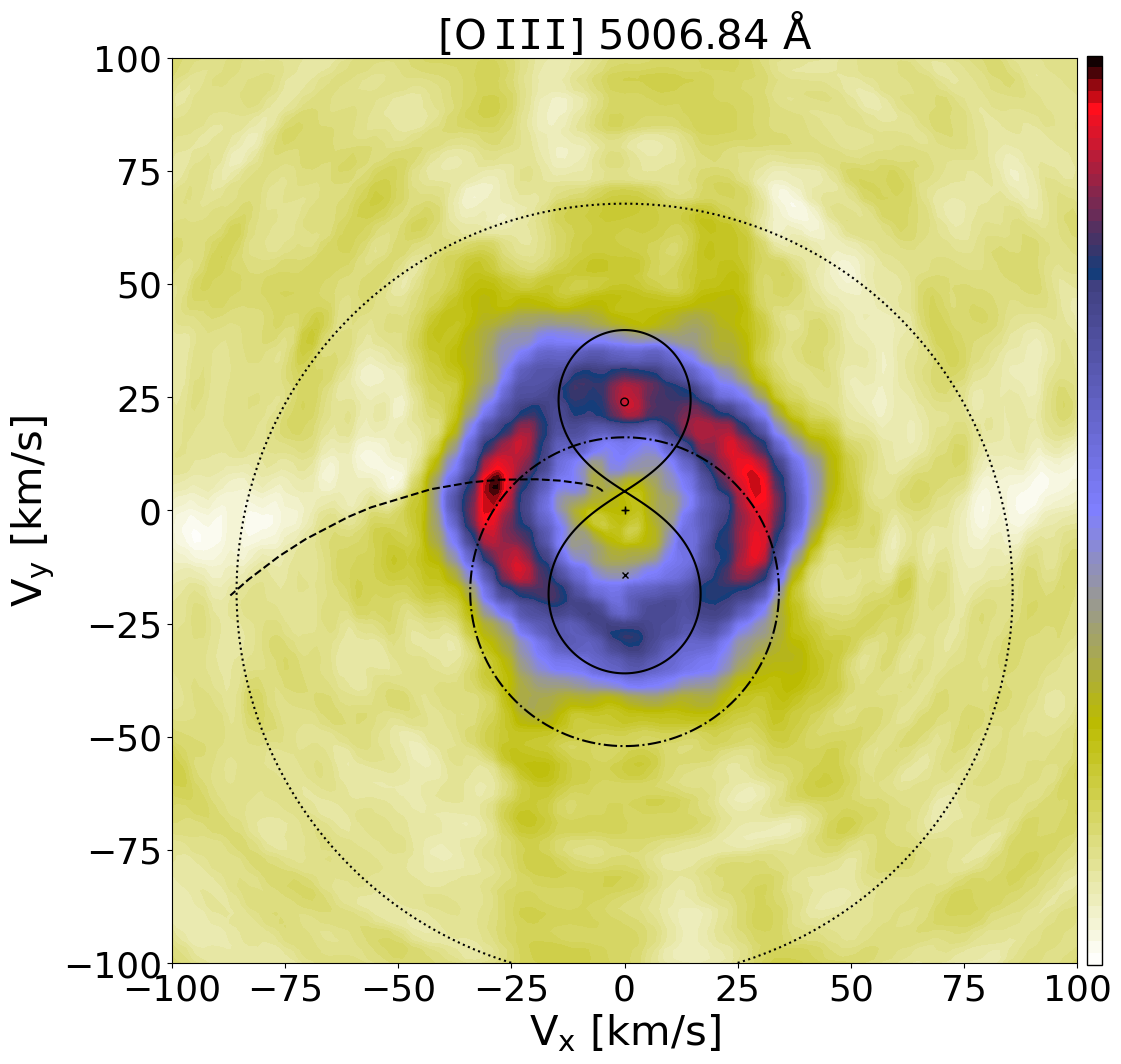}\hfill

    \includegraphics[width = 0.3\textwidth]{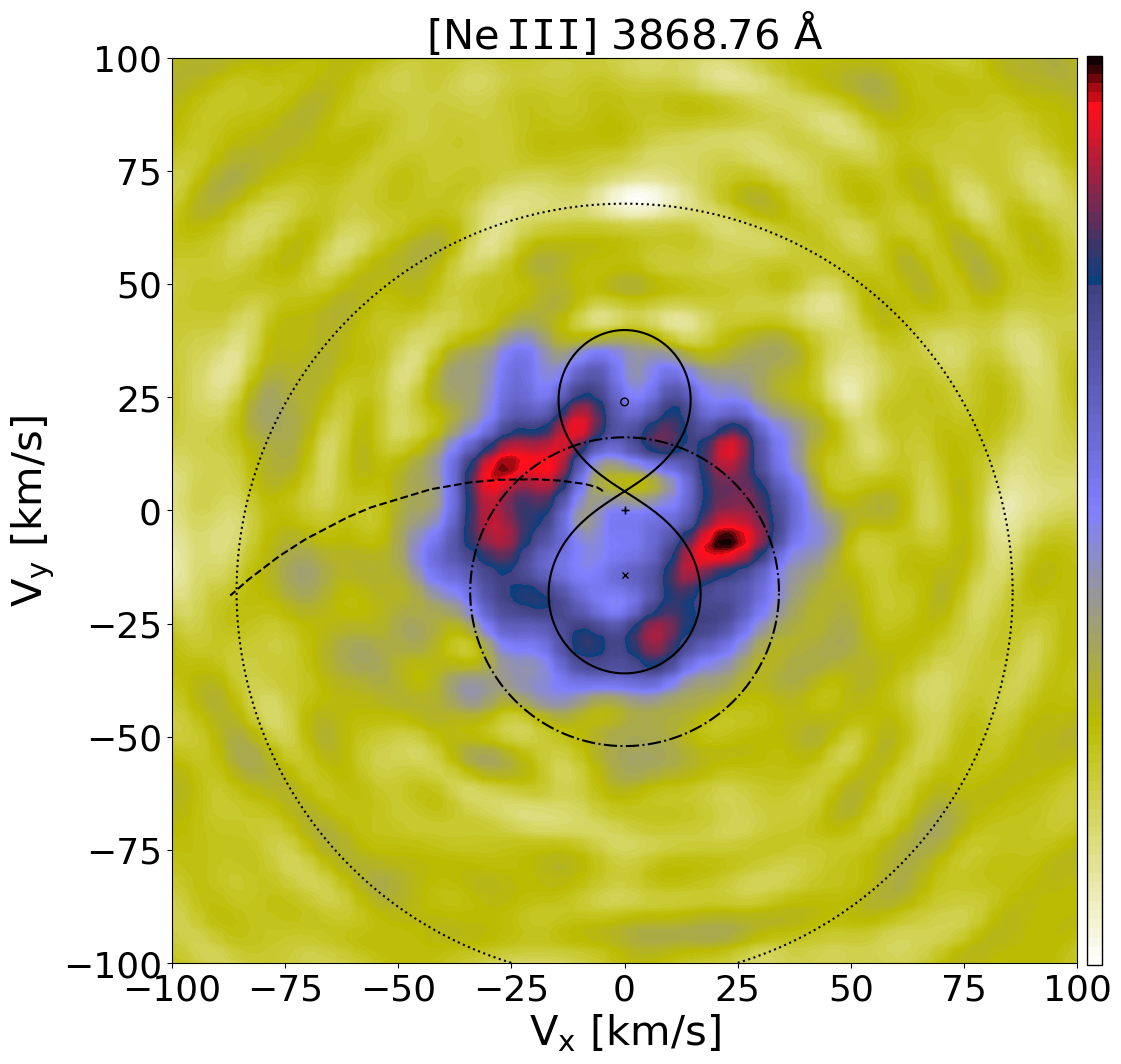}\hfill
    \includegraphics[width = 0.3\textwidth]{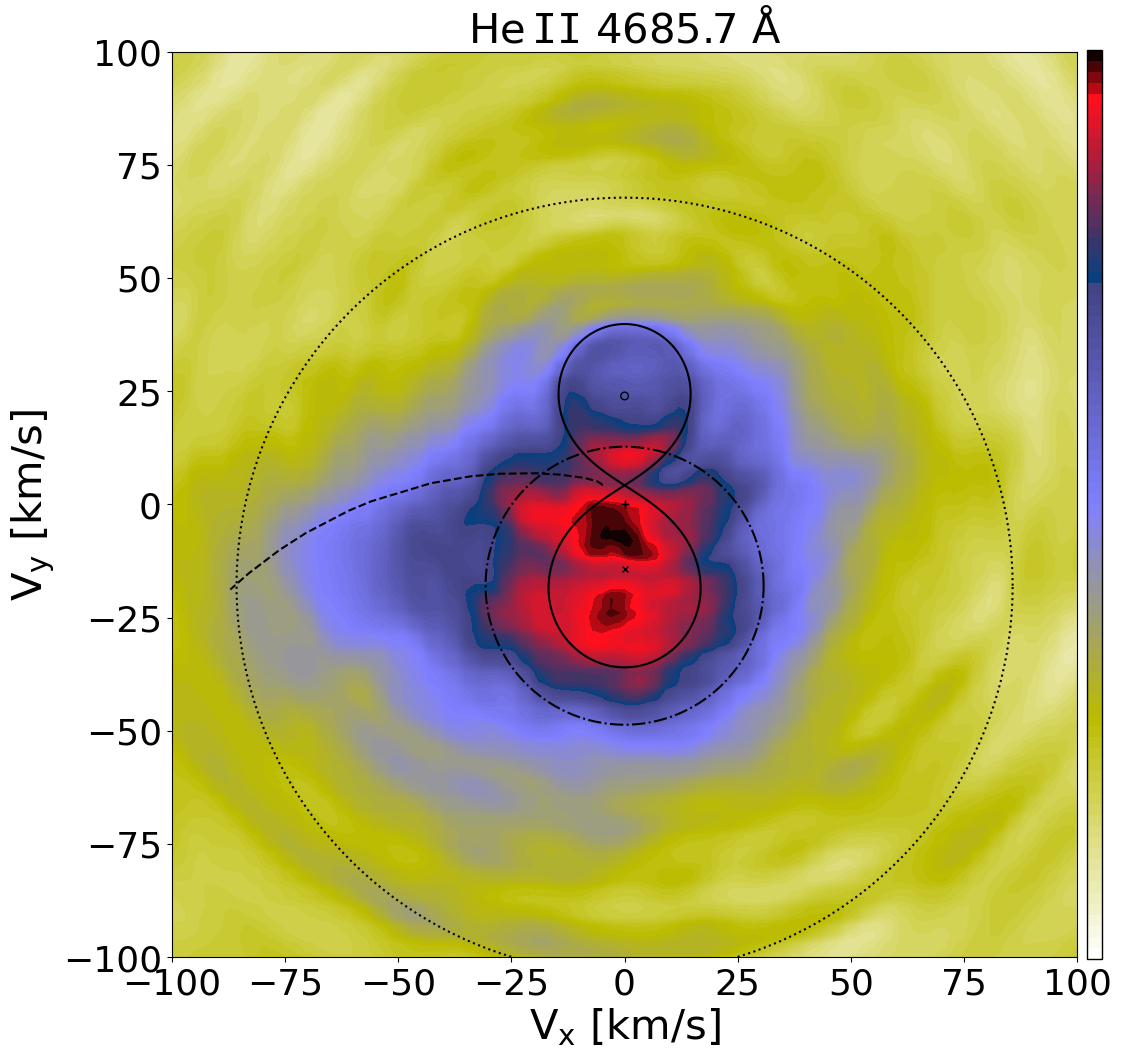}\hfill
    \includegraphics[width = 0.3\textwidth]{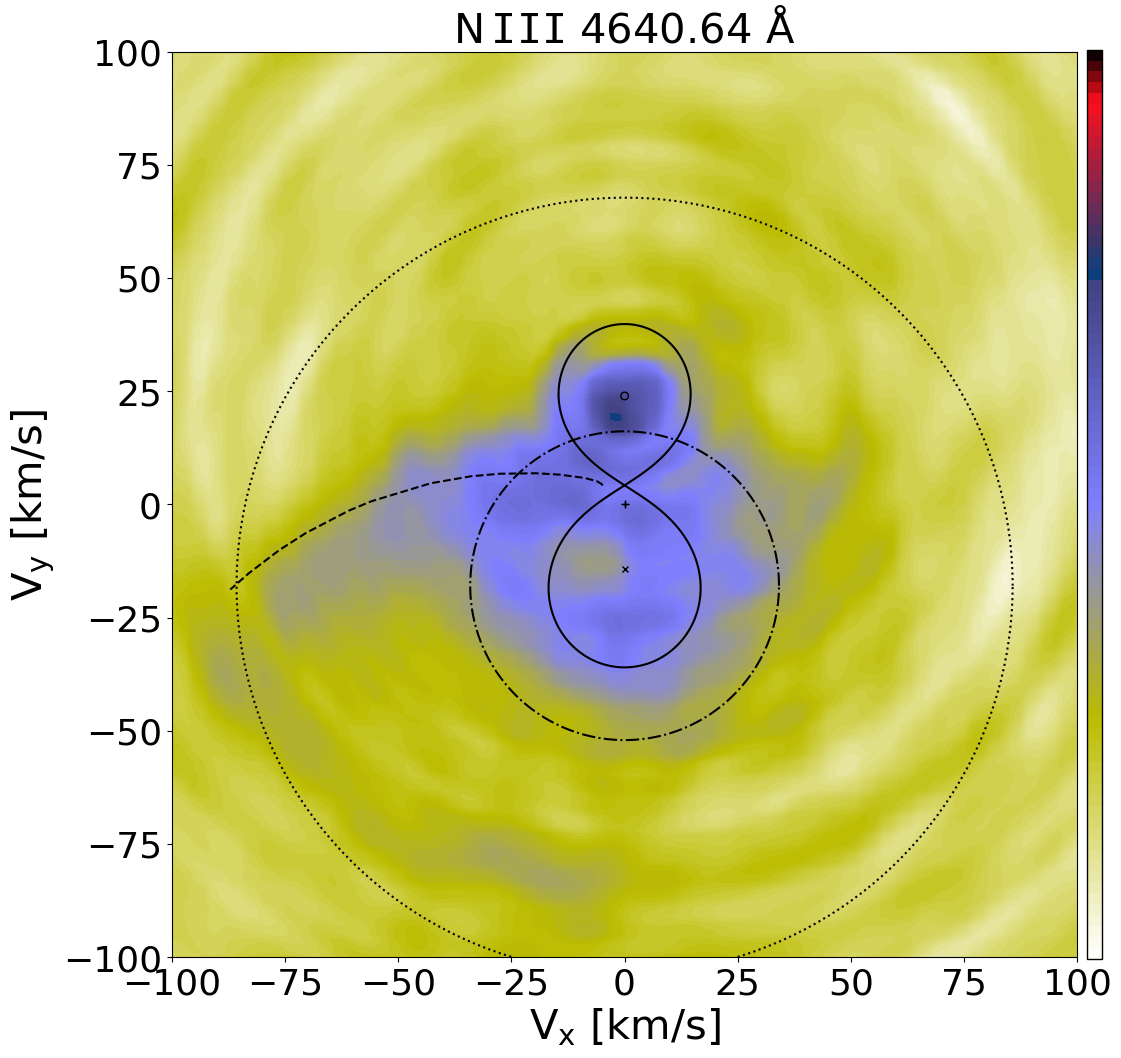}\hfill

    \caption{Doppler maps for the emission lines of Fig.~\ref{fig:spectra_interpol}}
    \label{fig:doppler_maps}
\end{figure*}

The dynamic spectra obtained during the SAP for twelve representative lines from Table~\ref{tab:spectral_lines} are displayed in Fig.~\ref{fig:spectra_interpol}. The color code is inverted for absorption lines (bottom row) to highlight the low flux values. 
The Doppler tomograms of the emission lines are displayed in Fig \ref{fig:doppler_maps}.

\FloatBarrier

\section{Secular evolution}
\begin{figure*}
    \includegraphics[width = 0.33\textwidth]{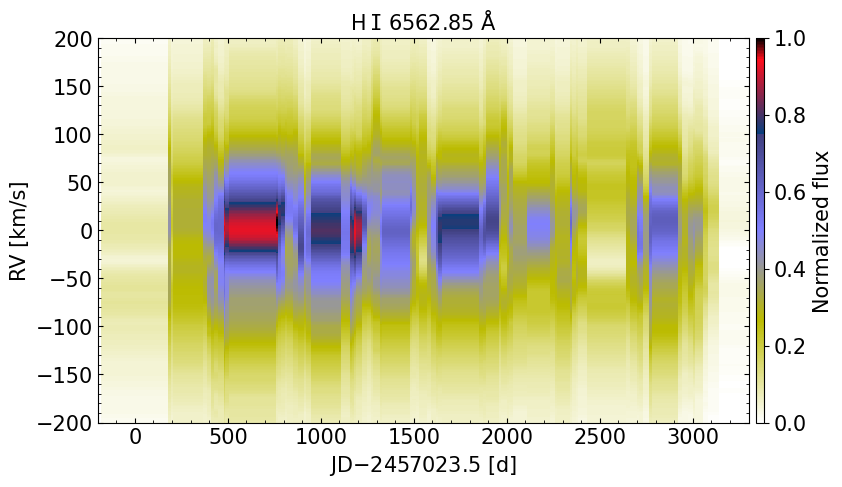}\hfill
    \includegraphics[width = 0.33\textwidth]{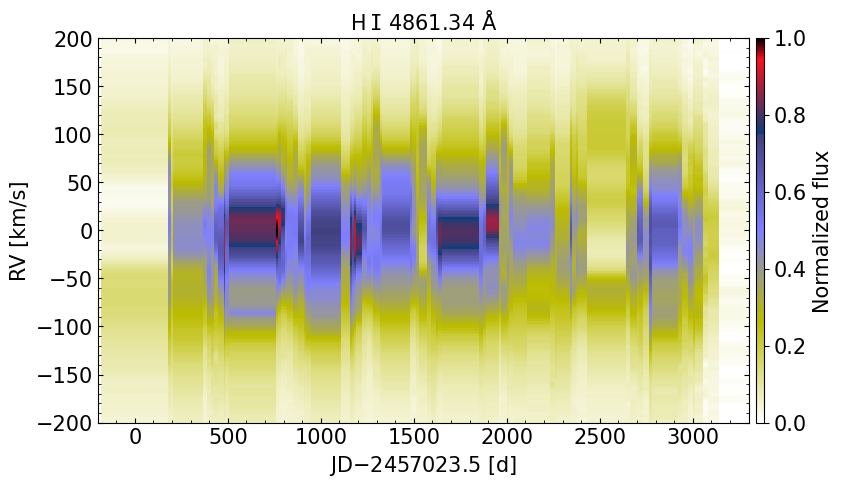}\hfill
     \includegraphics[width = 0.33\textwidth]{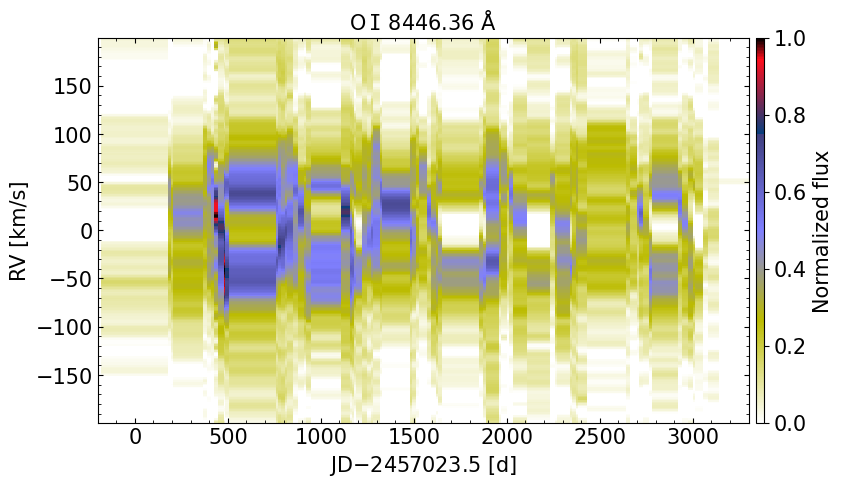}\hfill

    \includegraphics[width = 0.33\textwidth]{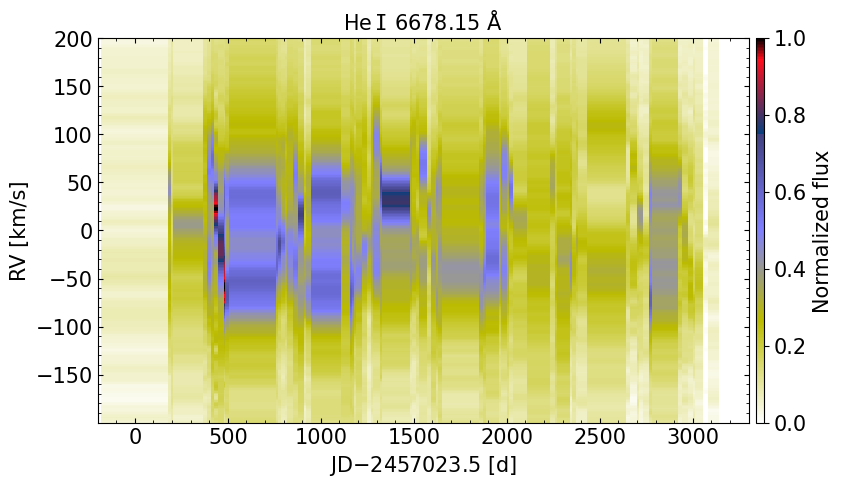}\hfill
    \includegraphics[width = 0.33\textwidth]{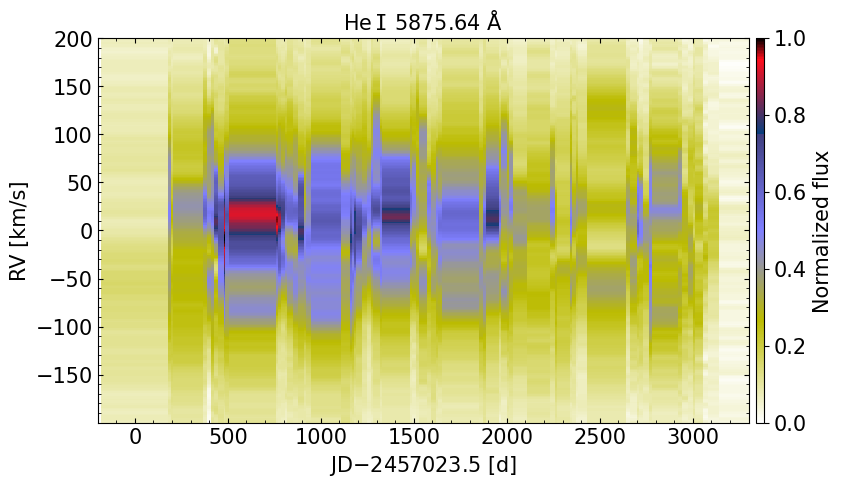}\hfill 
    \includegraphics[width = 0.33\textwidth]{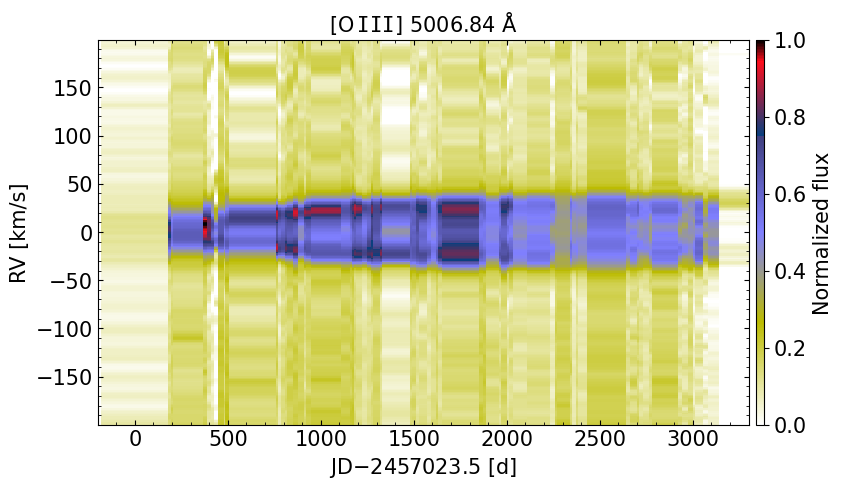}\hfill

    \includegraphics[width = 0.33\textwidth]{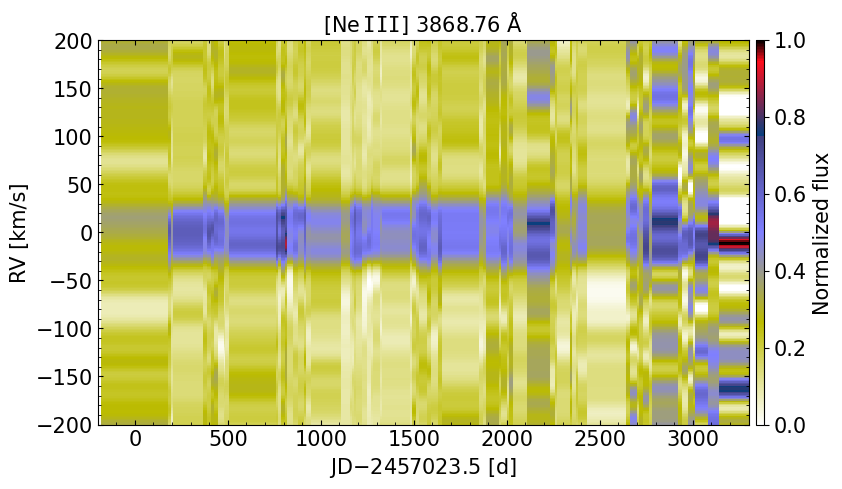}\hfill
    \includegraphics[width = 0.33\textwidth]{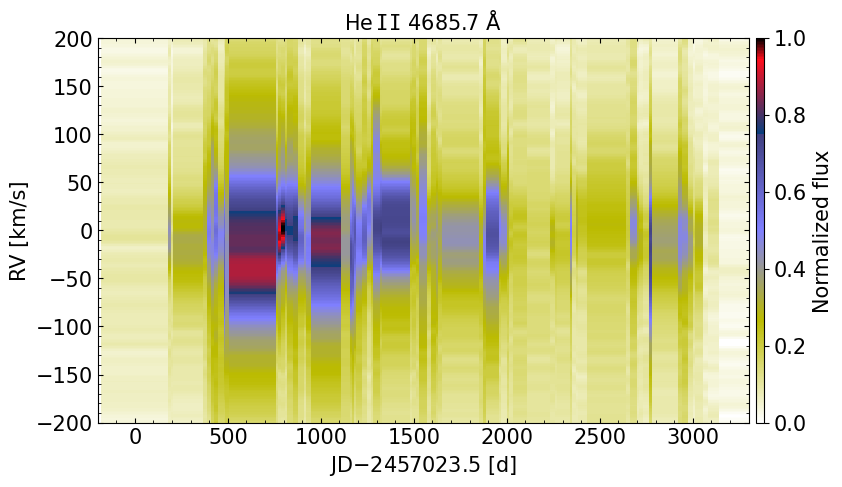}\hfill
    \includegraphics[width = 0.33\textwidth]{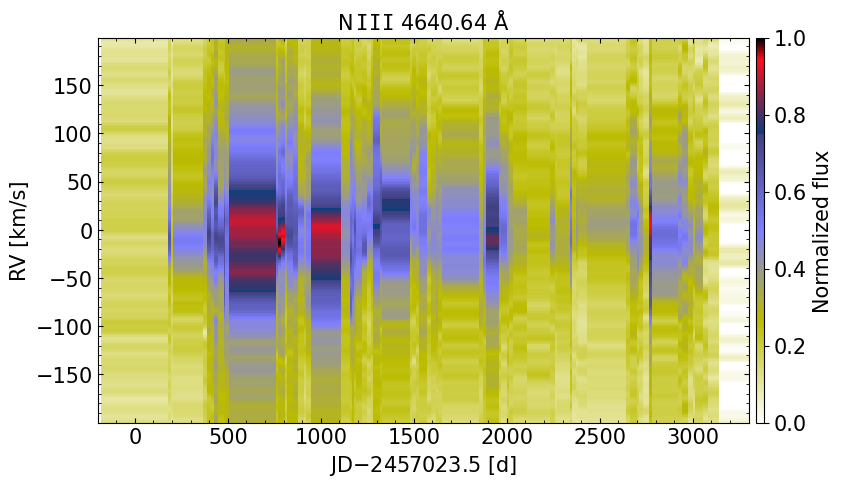}\hfill

    \caption{Temporal evolution of the emission lines of Fig.~\ref{fig:spectra_interpol}. Each panel shows the time elapsed since the SAP start ($t=0$),  with time running rightward.}
    \label{fig:timing_appendix}
\end{figure*}

Figure \ref{fig:timing_appendix} displays the temporal evolution of the emission component of the lines shown in Fig.~\ref{fig:spectra_interpol}. The start of the SAP, JD =2457023.5, defines $t=0$. The [\ion{Ne}{iii}] line, being at the border of the detector, is noise-dominated for times greater than 2800~days, when the blue continuum has dropped down. 

    

\end{appendix}
\end{document}